\documentclass[PAPER, atlasdraft=false, coverpage=false, cernpreprint=true, USenglish, texmf, orcidlogo]{atlasdoc}

\usepackage[subcaption]{atlaspackage}

\usepackage{dcolumn} %
\newcolumntype{d}[1]{D{.}{.}{#1}}
\newcolumntype{P}[1]{D{,}{\,\pm\,}{#1}}

\usepackage[acronym,shortcuts]{glossaries}
\glsdisablehyper

\usepackage{atlasbiblatex}
\usepackage{graphicx}
\usepackage{atlasphysics}
\graphicspath{{logos/}{figures/}}

\usepackage{ANA-SUSY-2022-11-PAPER-defs}

\AtlasTitle{Search for displaced leptons in $\sqrt{s}=13$~TeV and $13.6$~TeV $pp$ collisions with the ATLAS detector}

\AtlasVersion{2.13}

\AtlasAbstract{%

A search for leptons displaced from the primary vertex is performed with the ATLAS detector at the Large Hadron Collider. The search includes the full proton--proton collision dataset collected during Run 2 at $\sqrt{s}=13$~TeV and a partial dataset collected during Run 3 in 2022--2023 at
$\sqrt{s}=13.6$ TeV, corresponding to integrated luminosities of 140~fb$^{-1}$ and 56.3~fb$^{-1}$, respectively.
Final states with displaced electrons or muons are considered, and novel triggers introduced in Run 3 are employed that use large impact parameter tracking to reconstruct displaced tracks with low momentum.
In addition, photon reconstruction and multivariate techniques are employed to broaden the sensitivity to channels with large background rates or highly displaced electrons, respectively.
The results are consistent with the Standard Model background expectations and are used to set model-independent limits on the production of displaced electrons and muons.
The analysis is also interpreted in the context of a gauge-mediated supersymmetry breaking model with pair-produced long-lived sleptons, and a dark sector model with pair-produced chargino-like states.
The results include 95\% CL exclusions of selectrons with lifetimes from 4~ps to 60~ns and a mass of 150~GeV, and exclusions of selectrons, smuons, and staus with a lifetime of 0.3~ns for masses up to 740~GeV, 830~GeV, and 440~GeV, respectively.
Dark charginos with masses up to 380~GeV are excluded for a mass difference with the neutral state of 40~GeV, and mass differences down to 17~GeV are excluded for dark charginos with a 100~GeV mass.

}

\PreprintIdNumber{CERN-EP-2024-257}

\AtlasJournalRef{Phys.\ Rev.\ D 112 (2025) 012016}
\AtlasDOI{10.1103/w8hh-xf24}

\hypersetup{pdftitle={ATLAS document},pdfauthor={The ATLAS Collaboration}}

\begin{document}

\maketitle

\section{Introduction}
\label{sec:intro}


%
\Acp{LLP} occur in the \ac{SM}, and are predicted in a wide variety of \ac{BSM} scenarios, including both $R$-parity-conserving~\cite{Farrar:1978xj,Golfand:1971iw,Volkov:1973ix,Wess:1974tw,Wess:1974jb,Ferrara:1974pu,Salam:1974ig} and $R$-parity-violating supersymmetry (SUSY)~\cite{Barbier:2004ez,Allanach:2006st}, split-SUSY~\cite{Giudice:2004tc,ArkaniHamed:2004fb}, \ac{GMSB}~\cite{Dine:1981gu,AlvarezGaume:1981wy,Nappi:1982hm}, universal extra dimensions~\cite{Appelquist:2000nn,Cheng:2002ab}, and Hidden Valley models~\cite{Strassler:2006qa}.
Most searches for \ac{BSM} physics at the Large Hadron Collider (LHC)~\cite{Evans:2008zzb} experiments assume that BSM particles decay promptly, producing final state particles with trajectories consistent with an origin at the primary proton--proton (\pp) interaction point.
By contrast, LLPs with lifetimes longer than a few picoseconds travel at least hundreds of microns before decaying.
Their decay products can thus be displaced from the \ac{PV}, the vertex with the largest sum of squared track transverse momentum, or delayed with respect to the bunch crossing.
Reconstruction of the decay products can pose technical challenges that could have caused them to elude detection in prior searches.

\begin{figure}[b]
\centering
\begin{subfigure}{0.3\textwidth}
\centering
\includegraphics[width=\textwidth]{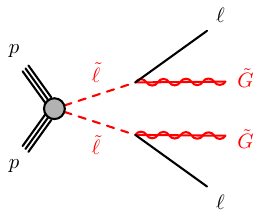}
\caption{}
\label{fig:signal_diagramsa}
\end{subfigure}
\begin{subfigure}{0.3\textwidth}
\centering
\includegraphics[width=\textwidth]{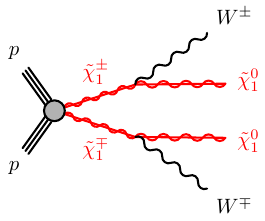}
\caption{}
\label{fig:signal_diagramsb}
\end{subfigure}
\caption{
Diagrams of the (a) GMSB slepton and (b) dark sector models.
}
\label{fig:signal_diagrams}
\end{figure}

This paper presents a search for \ac{BSM} physics producing events with two displaced light leptons (electrons or muons).
The search is performed with inclusive selection criteria to enable broad sensitivity to BSM models producing LLPs, with lifetimes from a few ps up to of order 100~ns, that decay into electrons or muons.
The sensitivity of the search is illustrated using a GMSB SUSY model~\cite{Alwall:2008ve,Alwall:2008ag,Alves:2011wf} in which the lightest SUSY particle (LSP) is a nearly massless gravitino ($\tilde{G}$) and the next-to-lightest SUSY particle (NLSP) is a slepton ($\tilde{\ell}$), which can be either a selectron ($\tilde{e}$), smuon ($\tilde{\mu}$), or stau ($\tilde{\tau}$).
The slepton can acquire a long lifetime due to its small gravitational coupling to the gravitino LSP.
As shown in Figure~\ref{fig:signal_diagrams}, the NLSP can be pair-produced, and then each decays into a charged SM lepton
($\ell$) of the same flavor and a gravitino LSP, yielding events containing two displaced leptons.
GMSB models often predict that staus are the NLSP~\cite{Evans_2016}.

Previous results setting limits on these processes
include an ATLAS search using the full Run~2 dataset, consisting of 140~fb$^{-1}$ of \pp collisions collected at $\sqrt{s} = 13~\TeV$, which excluded selectrons, smuons, and staus with masses up to 720~\GeV, 680~\GeV, and 340~\GeV, respectively, for a slepton lifetime of 0.1~ns~\cite{ATLAS:2020wjh}, dramatically enhancing the sensitivity with respect to prior constraints from the LEP experiments~\cite{Heister:2002vh,Abbiendi:2005gc,L3,Abdallah:2002rd,LEPlimitsSlepton}.
A CMS search based on 113--118~fb$^{-1}$ of $\sqrt{s} = 13~\TeV$ \pp collision data extended the stau mass reach to 405~\GeV for a similar lifetime~\cite{CMS-EXO-18-003}.
A subsequent full Run~2 ATLAS search for muons with intermediate displacements, in between those of the displaced and prompt lepton searches, extended the sensitivity to GMSB smuons with lifetimes below about 0.01 ns~\cite{SUSY-2020-09}.
Additional searches for disappearing track signatures in Run~2 data by ATLAS~\cite{ATLAS:2022rme} and CMS~\cite{CMS:2023mny}, as well as a Run~2 ATLAS search for tracks with high ionization energy loss $dE/dx$~\cite{ATLAS:2022pib} could also be sensitive to this GMSB slepton model for various lifetimes.

Interpretations are also performed using a Higgs portal model that introduces a dark sector consisting of a
charged state ($\tilde{\chi}^\pm$) and neutral state ($\tilde{\chi}^0$) with mass difference
$\Delta m \equiv m(\tilde{\chi}^\pm)-m(\tilde{\chi}^0)$~\cite{charginos1},
which can produce the observed dark matter relic density for $10 \lesssim \Delta m \lesssim 40$~\GeV~\cite{charginos2}.
For the signal process $\tilde{\chi}^+\tilde{\chi}^-\rightarrow W^+ W^- \tilde{\chi}^0 \tilde{\chi}^0$
shown in Figure~\ref{fig:signal_diagrams}, the charged state can acquire a non-negligible lifetime due to the weak coupling between the dark states and the $W$ boson.
This results in events with two displaced leptons if the $W$ bosons decay leptonically, the same final state as with the GMSB slepton model.
However, with this model the leptons tend to have lower \pt since the lighter neutral state is massive, highlighting the need for improved low-momentum lepton acceptance.
In SUSY models, as assumed here, the chargino and neutralino play the roles of the new charged and neutral states, respectively,
and chargino-pair production cross-section is used for the signal model.

This result enhances the discovery reach for displaced leptons beyond that of prior searches through several novel additions.
It includes data collected by the ATLAS detector at a center-of-mass energy of 13.6~TeV, new triggers introduced in Run~3 that exploit improved tracking capabilities of the \ac{HLT} to enable lower lepton transverse momentum (\pT) thresholds, as well as improved offline tracking performance that greatly lowers the rate of incorrectly reconstructed displaced tracks~\cite{IDTR-2022-04}.
Precision information from the ATLAS \ac{LAr} \ac{EM} calorimeter is introduced for sensitivity to highly displaced electrons, which may fail to include reconstructed tracks and be reconstructed as photons.
Finally, the analysis includes a first consideration of single displaced electron channels, which were previously inaccessible due to high background rates and are enabled using multivariate \acp{BDT}.

The analysis combines two approaches with mutually exclusive selections: the ``ABCD'' search and the ``EM-BDT'' search.
In both cases, backgrounds are determined with entirely data-driven methods, and \ac{MC} simulation is only used for signal models.
The ABCD search targets dilepton events, requiring at least two electrons, two muons, or an electron and a muon, and uses similar techniques as the prior Run~2 ATLAS displaced leptons search~\cite{ATLAS:2020wjh} for the estimation of backgrounds from \ac{FHF} leptons and cosmic-ray muons.
The ABCD search is performed in two orthogonal sets of \acp{SR} with different triggers and lepton \pt thresholds.
The ``high-\pt'' SRs use a similar selection as the prior Run~2 search~\cite{ATLAS:2020wjh}, %
and the new \ac{LRT} SRs use novel Run~3 triggers that employ \ac{LRT}~\cite{TRIG-2022-01} (see Section~\ref{sec:LRT}) to extend the trigger tracking to large transverse impact parameter ($|\dzero|$) values, thus  enabling reduced lepton thresholds with respect to Run~2. %
A separate EM-BDT search also exploits the new LRT trigger for electrons, and leverages precision tracking and electromagnetic calorimeter information (see Section~\ref{sec:LAr}) to
improve the displaced e/$\gamma$ selection, building on previous ATLAS Run~2 searches for displaced photons~\cite{SUSY-2020-28,ATLAS:2022vhr}.
The LRT channel of the ABCD search and the EM-BDT search both use 56.3~\ifb\ of Run~3 data collected at $\sqrt{s}=13.6$~\TeV in 2022--2023, while the ABCD high-\pt channel also uses the full Run~2 dataset of 140~\ifb\ collected at $\sqrt{s}=13$~\TeV.

The paper is organized as follows.
Section~\ref{sec:detector} describes the ATLAS detector, with specific emphasis on the unique \ac{LRT} and LAr reconstruction techniques employed in the analysis.
Section~\ref{sec:datamc} describes the data and simulated samples used, and selections for objects and analysis events are given in Section~\ref{sec:objects}.
Descriptions of the ABCD (Section~\ref{sec:abcd}) and EM-BDT (Section~\ref{sec:embdt}) analyses follow, including detailed event selections, background estimations, and systematic uncertainties.
Finally, results and interpretations are presented in Section~\ref{sec:result}, followed by brief conclusions.


%
\FloatBarrier
\section{ATLAS detector}
\label{sec:detector}
%

%
The ATLAS detector~\cite{PERF-2007-01,GENR-2019-02} at the LHC covers nearly the entire solid angle around the collision point.\footnote{ATLAS uses a right-handed coordinate system with its origin at the nominal interaction point (IP)
in the center of the detector and the \(z\)-axis along the beam pipe.
The \(x\)-axis points from the IP to the center of the LHC ring,
and the \(y\)-axis points upwards.
Polar coordinates \((r,\phi)\) are used in the transverse plane,
\(\phi\) being the azimuthal angle around the \(z\)-axis.
The pseudorapidity is defined in terms of the polar angle \(\theta\) as \(\eta = -\ln \tan(\theta/2)\) and is equal to the rapidity
$ y = \frac{1}{2} \ln \left( \frac{E + p_z}{E - p_z} \right) $ in the relativistic limit.
Angular distance is measured in units of \(\Delta R \equiv \sqrt{(\Delta y)^{2} + (\Delta\phi)^{2}}\).}
It consists of an inner tracking detector surrounded by a thin superconducting solenoid, electromagnetic and hadronic calorimeters,
and a muon spectrometer incorporating three large superconducting air-core toroidal magnet systems.

\subsection{Detector subsystems}

The inner-detector system (ID) is immersed in a \qty{2}{\tesla} axial magnetic field
and provides charged-particle tracking in the range \(|\eta| < 2.5\).
The high-granularity silicon pixel detector covers the interaction region and typically provides four measurements per track,
the first hit generally being in the insertable B-layer (IBL). %
It is followed by the SemiConductor Tracker (SCT), which usually provides eight measurements per track.
These silicon detectors are complemented by the transition radiation tracker (TRT),
which enables radially extended track reconstruction up to \(|\eta| = 2.0\).
The TRT also provides electron identification information
based on the fraction of hits (typically 30 in total) above a higher energy-deposit threshold corresponding to transition radiation.

The calorimeter system covers the pseudorapidity range \(|\eta| < 4.9\).
Within the region \(|\eta|< 3.2\), \ac{EM} calorimetry is provided by barrel and
endcap high-granularity lead/LAr calorimeters,
with an additional thin LAr presampler covering \(|\eta| < 1.8\)
to correct for energy loss in material upstream of the calorimeters.
Hadronic calorimetry is provided by the steel/scintillator-tile calorimeter,
segmented into three barrel structures within \(|\eta| < 1.7\), and two copper/LAr hadronic endcap calorimeters.
The solid angle coverage is completed with forward copper/LAr and tungsten/LAr calorimeter modules
optimized for electromagnetic and hadronic energy measurements, respectively.

The muon spectrometer (MS) comprises separate trigger and
high-precision tracking chambers measuring the deflection of muons in a magnetic field generated by the superconducting air-core toroidal magnets.
The field integral of the toroids ranges between \num{2.0} and \qty{6.0}{\tesla\metre}
across most of the detector.
Three layers of precision chambers, each consisting of layers of monitored drift tubes, cover the region \(|\eta| < 2.7\),
except in the innermost layer of the endcap region, where layers of small-strip thin-gap chambers and Micromegas chambers both provide precision tracking in the region \(1.3 < |\eta| < 2.7\).
The muon trigger system covers the range \(|\eta| < 2.4\) with resistive-plate chambers in the barrel, thin-gap chambers in the endcap regions, and the aforementioned small-strip thin-gap chambers and Micromegas chambers in the innermost layer of the endcap.
These replaced cathode-strip chambers that were used through Run~2.

The luminosity is measured mainly by the LUCID--2 detector that records Cherenkov light produced in the quartz windows of photomultipliers located close to the beampipe.

Events are selected by the first-level trigger system implemented in custom hardware,
followed by selections made by algorithms implemented in software in the \ac{HLT}~\cite{TRIG-2016-01,TRIG-2022-01}.
The first-level trigger accepts events from the \qty{40}{\MHz} bunch crossings at a rate below \qty{100}{\kHz},
which the high-level trigger further reduces in order to record complete events to disk.

The Run~3 detector configuration benefits from several upgrades compared with that of Run~2 to maintain high detector performance at the higher pileup levels of Run~3. The improvements include a new innermost layer of the muon spectrometer in the endcap region, which provides higher redundancy and a large reduction in fake muon triggers. The trigger system also benefits from new LAr digital electronics with significantly increased granularity. Other updates and further details are provided in Ref.~\cite{GENR-2019-02}.
A software suite~\cite{SOFT-2022-02} is used in data simulation, in the reconstruction
and analysis of real and simulated data, in detector operations, and in the trigger and data acquisition
systems of the experiment.

As described in more detail below, this analysis
exploits some specialized capabilities of the ATLAS detector
and its TDAQ system.
These include \ac{LRT}
(see Section~\ref{sec:LRT}), which is the ability to reconstruct tracks with a large
impact parameter with respect to the PV, and the
ability of the EM calorimeter to measure the
flight direction and time-of-arrival of particles
causing EM showers (see Section~\ref{sec:LAr}).

\subsection{Large radius tracking}
\label{sec:LRT}

The standard ATLAS tracking algorithm is designed to reconstruct the trajectory of promptly produced charged particles.
It is efficient out to a $|\dzero|$ of 5~mm, limiting its usefulness for LLPs that decay into charged particles after traversing part of the detector.
The \Ac{LRT}~\cite{IDTR-2021-03} is a configuration of the tracking algorithm designed to increase the tracking efficiency for decay products of LLPs.
Figure~\ref{fig:timing_d0a} shows an example $|\dzero|$ distribution, extending well past 5~mm, for several models in the EM-BDT search, which combines both standard and \ac{LRT} tracking.
For the offline reconstruction of data, it is implemented as a second pass, running on the remaining hits that were not used by the standard tracking, with a looser restriction on the impact parameters and other optimizations, e.g.\ in the seeding step.
Large radius tracks are reconstructed out to a $|\dzero|$ of 300~mm, the first layer of the SCT.
Compared to previous searches during Run 2, the \ac{LRT} configuration in the reconstruction software was improved to greatly reduce the number of incorrectly reconstructed tracks as well as to reduce the computation time, allowing \ac{LRT} tracks to be reconstructed in all events.
This improved reconstruction is used consistently for both the Run~2 and Run~3 data where the performance is similar since the inner detector geometry is the same for both runs.
For the reconstruction of offline electrons and muons, described in Section~\ref{sec:objects}, both standard and LRT tracks are used.

In Run~3, \ac{LRT} is run in the \ac{HLT} for the first time~\cite{TRIG-2022-01}.
For signal-like electrons and muons, the \ac{LRT} used in the \ac{HLT} is expected to be more than 80\% efficient with respect to offline reconstruction for $|d_0|<125$~mm.
This is possible due to improvements in the LRT computation time, the HLT tracking algorithms, and the HLT computing resources.
\ac{LRT} for leptons in the trigger is implemented as a single pass, running on all hits in small regions around lepton candidates from the first-level trigger, with the same looser restrictions on impact parameters and similar optimizations as the offline reconstruction.
This means the LRT-based lepton triggers do not use standard tracking.
The same first-level hardware triggers for the primary prompt electron and muon triggers are used.
The total trigger efficiency depends on the lepton flavor and kinematics \cite{TRIG-2022-01}.
For example, for mouns it drops from 90\% at small production radius to 60\% at 300~mm.

\begin{figure}[!t]
\centering
\begin{subfigure}{0.49\textwidth}
\centering
\includegraphics[width=\textwidth]{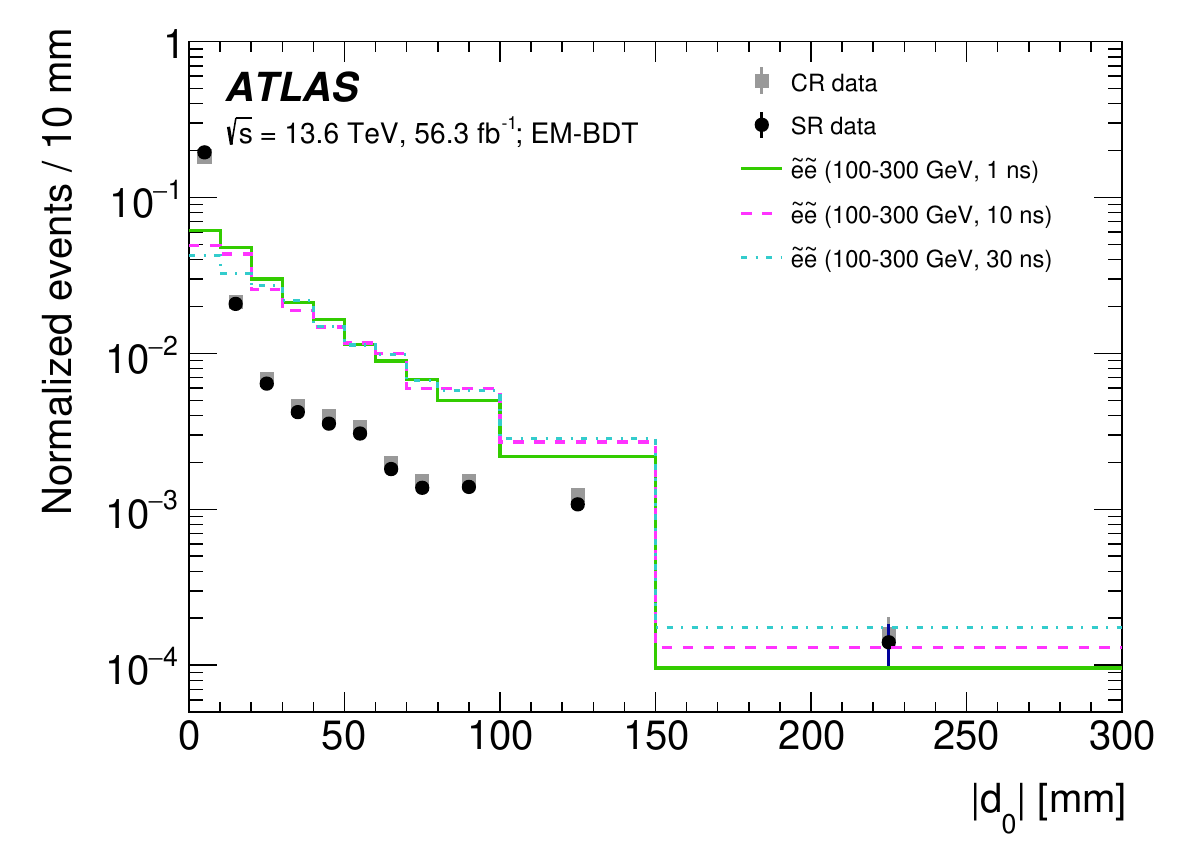}
\caption{}
\label{fig:timing_d0a}
\end{subfigure}
\begin{subfigure}{0.49\textwidth}
\centering
\includegraphics[width=\textwidth]{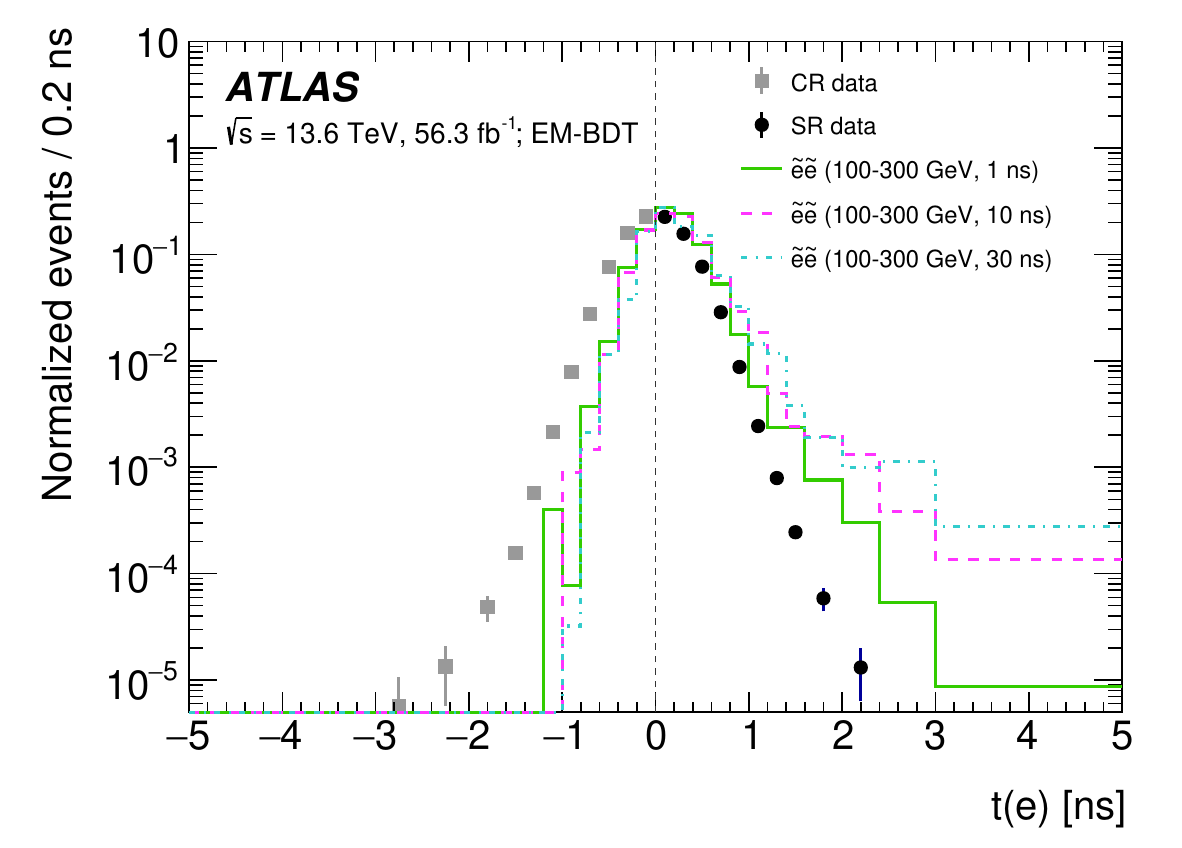}
\caption{}
\label{fig:timing_d0b}
\end{subfigure}
\caption{Distributions of the (a) $|d_{0}|$ and (b) LAr timing using the EM-BDT search regions with Run 3 data and selected signal samples with slepton masses of 100, 200, and 300~\GeV.
The LAr timing measurement was calibrated to zero for promptly produced particles traveling near the speed of light.
The data events with negative times (gray squares) represent the
signal-depleted control region (CR) used to determine the
background estimate, while the data events with positive times (black circles)
represent the signal region (SR) events on which the final search is performed. The SR data were blinded until the analysis strategy was finalized and the data were unblinded to determine the results.}
\label{fig:timing_d0}
\end{figure}

\subsection{Electromagnetic calorimeter timing and pointing measurements}
\label{sec:LAr}

Electrons from the decays of massive LLPs could reach the LAr calorimeter with a slight delay compared to prompt electrons. This delay arises mostly from the LLP time-of-flight,
which corresponds to a relativistic speed ($\beta = v/c$) that can be substantially less than one.

In addition, the opening angle in the LLP decay causes the electron to be ``non-pointing'', arriving at the calorimeter from a direction that does not point directly back to the PV.
This effect results in the geometrical path to the calorimeter being longer than that for a prompt electron from the PV, further increasing the delay in its arrival time.
Figure~\ref{fig:timing_d0b} illustrates the delay in the timing distribution of several signal models in the EM-BDT SR.

As described in Refs.~\cite{SUSY-2020-28,ATLAS:2022vhr} and references therein, the fine segmentation of the LAr EM calorimeter can be used to measure the flight direction of a particle (electron or photon) that produces an EM shower, while its accordion structure and fast readout allows precise measurement of the particle's arrival time at the calorimeter.
A line drawn through the centroids of the EM shower in the first two longitudinal layers is used to calculate the pointing value, defined as the difference between the position on the $z$-axis where
the line crosses the beamline and the position of the PV.
The resolution on the pointing measurement is $\approx$~15~mm for an EM shower with energy of $\approx$~50--100~\GeV in the barrel.
The arrival time is determined using the energy deposit from the second-layer EM calorimeter cell with the maximum energy deposit among cells in the associated EM shower (\maxEcell).
For an electron with an energy in the range of interest, this cell typically contains about 20\%--50\% of the total energy deposited in the EM shower.
The timing resolution is approximately 200~ps for EM showers with \maxEcell\ greater than 20~\GeV, and is dominated by the LHC beamspread.


%
%

\FloatBarrier
\section{Data and Simulation samples}
\label{sec:datamc}

Collision data collected by the ATLAS detector from the LHC's $\sqrt{s} = 13~\TeV$ $pp$  Run~2 (2015--2018) and  $\sqrt{s} = 13.6~\TeV$ $pp$  Run~3 (2022--2023) are used,
with luminosities corresponding to $140~\ifb$ and $56.3~\ifb$, respectively, after applying
detector conditions and data quality requirements~\cite{DAPR-2018-01}.
The uncertainty in the Run~2 integrated luminosity is 0.83\%~\cite{DAPR-2021-01}, for 2022 it is 2.2\%~\cite{ATL-DAPR-PUB-2023-001}, and for 2023 it is 2.0\%~\cite{ATL-DAPR-PUB-2024-001}, following the methodology discussed in Ref.~\cite{DAPR-2021-01}, using the LUCID--2 detector~\cite{LUCID2} for the primary luminosity measurements, complemented by measurements using the inner detector and calorimeters.
The ABCD high-\pt search uses the Run 2 and Run 3 datasets.
The ABCD LRT and EM-BDT searches use only the Run 3 data.
Run~2 data was reconstructed with a consistent software suite~\cite{SOFT-2022-02} as used for Run~3 data to take advantage of developments, such as the improved LRT described in Section~\ref{sec:LRT}.

Data were collected with several unprescaled triggers~\cite{TRIG-2022-01,TRIG-2016-01} with signatures and thresholds that would accept events from displaced decays of leptons.
Electron~\cite{TRIG-2018-05} and muon~\cite{TRIG-2018-01} triggers based on standard tracking~\cite{TRIG-2019-03} are efficient out to $|\dzero|$ values of approximately 5 or 10~mm, respectively.
Thus they are not optimal for saving events with LLPs of moderate lifetime.
Photon triggers accept events with electrons, even if they are too displaced to reconstruct a track, but have higher energy thresholds than electron triggers.
MS-only triggers accept events based on muon tracks in the MS, without the need for an inner-detector track; these also have higher thresholds and some have restrictions in $\eta$ to additionally reduce the trigger rate.
Thanks to the requirement of an inner detector track, LRT-based lepton triggers are able to save events with lower momentum thresholds of 30 and 20~\GeV for electrons and muons, with $|\dzero|$ thresholds of 3 and 2~mm, respectively.
Table~\ref{tab:triggers} lists the triggers used based on the analysis region and lepton topology.

\begin{table}[!t]
\small
\begin{center}
\caption{Summary of trigger thresholds used for the ABCD and EM-BDT analyses, based on the object topology.
Trigger objects are denoted as $\gamma$ for photons, $e$ for electrons, and $\mu$ for muons, preceded by the multiplicity and followed by the \pT threshold in~\GeV.
``MS-only'' indicates triggers that only rely on the Muon Spectrometer (MS), and not the inner detector.
For example, $2\gamma50$ means a trigger requiring the presence of two photons with $\pt>50~\gev$.
The $1\mu60$ MS-only trigger has a restriction of $|\eta| < 1.05$.
The \acrfull{LRT} electron (muon) trigger has a $|\dzero| > 3(2)$~mm requirement.
None of the triggers have an isolation requirement.
Triggers listed with `Run~3 only' were only used for the corresponding analysis region to collect data in Run~3, otherwise they were also used for Run~2 data.
``OR'' indicates a logical OR of multiple triggers.}
\begin{tabular}{ll}
\noalign{\smallskip}\hline\noalign{\smallskip}
Topology & Trigger Thresholds [GeV] \\
\noalign{\smallskip}\hline\noalign{\smallskip}
\multicolumn{2}{c}{ABCD high-\pt} \\
\noalign{\smallskip}\hline\noalign{\smallskip}
\texttt{if} $\geq 1e,~\pt > 160~\gev$                                        &  1$\gamma 140$ \\
\texttt{else if} $\geq 2e,~\pt > 60~\gev$                                   &  $2\gamma50$ \\
\texttt{else if} $\geq 1 \mu,~\pt > 60~\gev,~|\eta| < 1.07$                 &  $1\mu60$ (MS-only)  \\
\texttt{else if} $e\mu$  (Run 3 only)   &  $\gamma40\mu40$ (MS-only)  \\
\texttt{else if} $\mu\mu$ (Run 3 only)  &  $2\mu50$ (MS-only)  \\
\noalign{\smallskip}\hline\noalign{\smallskip}
\multicolumn{2}{c}{ABCD LRT}  \\
\noalign{\smallskip}\hline\noalign{\smallskip}
$e, \pt > 31$~\GeV (Run 3 only)                  &  $1e30$ (LRT) \\
$\mu, \pt > 21$~\GeV  (Run 3 only)               &  $1\mu20$ (LRT)\\
\noalign{\smallskip}\hline\noalign{\smallskip}
\multicolumn{2}{c}{EM-BDT}  \\
\noalign{\smallskip}\hline\noalign{\smallskip}
$=1e$, $0\gamma$ (Run 3 only)        &  $1e30$ (LRT) OR $1\gamma140$     \\
$\geq2e$ OR ($1e, 1\gamma$) (Run 3 only) & $1e30$ (LRT) OR $2\gamma50$ OR  $1\gamma140$  \\
$\geq2\gamma$ (Run 3 only)         &  $2\gamma50$ OR $1\gamma140$ \\
\noalign{\smallskip}\hline\noalign{\smallskip}
\end{tabular}
\label{tab:triggers}
\end{center}
\end{table}

For the baseline GMSB-motivated
slepton signal model in Figure~\ref{fig:signal_diagrams},
\ac{MC} simulation events are used to determine the lepton selection criteria and estimate the sensitivity
in a simplified GMSB SUSY model with light sleptons and a gravitino LSP with mass much less than 1~\GeV.
The signal models are given by the slepton flavor, mass, and lifetime.
For each slepton mass value, several slepton
lifetimes were simulated, typically ranging from 1~ps to 30~ns.
Since the distribution of particle flight distances follows an exponential
distribution,
it is possible to reweight the shape of that curve and thus
generalize from the simulated samples to other lifetime values.
Each event is assigned a weight according to a source signal lifetime, target signal lifetime, and the decay of the event in question.

For the dark sector model in Figure~\ref{fig:signal_diagramsb}, \ac{MC} simulation is used to perform two scans of parameter space: varying the mass difference $\Delta m$ and chargino mass for a fixed chargino lifetime of  $\tau=0.3$~ns, and varying the chargino lifetime and mass for a fixed $\Delta m=40$~\GeV.
This choice was motivated by the analysis having the best expected sensitivity to the GMSB model for a lifetime of 0.3~ns, and a mass splitting of 40~\GeV is the most favorable in terms of lepton \pt that is still consistent with the dark matter relic density.
The $W$ bosons are forced to decay into an electron or muon (no taus) with equal probability in the simulation.

Signal model MC simulation events with up to two additional partons at leading order are simulated using \MGMCatNLOV{v2.6.1}~\cite{Alwall:2014hca} interfaced to \PYTHIAV{8.230}~\cite{pythia8ref} with the A14 parameter set (tune)~\cite{ATL-PHYS-PUB-2014-021} along with the \nnpdftwo parton distribution function (PDF) set~\cite{Ball:2012cx}.
$\GEANT$~\cite{Agostinelli:2002hh} is used to simulate the propagation and decay of the slepton, but without preserving information about its chirality.
To model the impact of multiple interactions that occur
in the same and neighboring
bunch crossings (pileup), each hard-scattering event
is overlaid~\cite{SIMU-2020-01}
with simulated minimum bias events generated by EPOS 2.0.1.4~\cite{Porteboeuf:2010um,Pierog:2013ria} with the EPOS LHC tune
and \PYTHIAV{8.307}~\cite{pythia83} with the \nnpdftwo set of PDFs and A3
tune~\cite{ATL-PHYS-PUB-2016-017}.
The 13~\TeV\ wino-like chargino-pair and slepton-pair production cross-sections and uncertainties were calculated at
next-to-leading-order (NLO) in $\alpha_\text{s}$, with soft-gluon emission effects at next-to-leading-logarithm~\cite{Beenakker:1999xh,Debove:2010kf,Fuks:2012qx,Fuks:2013vua,Fiaschi:2018hgm,Bozzi:2007qr,Fuks:2013lya,Fiaschi:2018xdm}.
At 13.6~\TeV, these were calculated at approximate next-to-next-to-leading-order in $\alpha_\text{s}$ matched with threshold resummation at the next-to-next-to-leading-logarithmic accuracy (aNNLO+NNLL) using an envelope of predictions with different factorization and renormalization scales and PDF sets~\cite{Fiaschi:2023tkq}.
The superpartners of the left- and right-handed leptons are assumed to have equal mass.
The slepton-pair cross-section for a single flavor of $\tilde{\ell}$
with mass $100~\GeV$ ($800~\GeV$)
is $0.37 \pm 0.01$~pb ($0.059 \pm 0.004$~fb) at 13~TeV and $0.41 \pm 0.01$~pb ($0.069 \pm 0.008$~fb) at 13.6~TeV.
The chargino-pair cross-section for a mass of 300~\GeV is $190\pm13$~fb at 13~\TeV and $216\pm9$~fb at 13.6~\TeV.
A factor of 0.0642 is included to account for the branching ratio of the two W boson decays into electrons and muons, which accounts for decays via taus.


%
\FloatBarrier
\section{Object selections and event preselections}
\label{sec:objects}

%
%
%
%
%
%

The main objects used in this analysis are electrons and muons, including those produced in the decays of $\tau$-leptons, and photons. Jets, including those from hadronically decaying $\tau$-leptons, are also used in the procedure to remove ambiguities between nearby reconstructed objects (``overlap removal'') and to construct the magnitude of the missing transverse momentum (\MET), which is used in the EM-BDT analysis.

For electrons, muons, and photons, a baseline selection is used for the overlap removal procedure.
For electrons and muons, additional kinematic requirements are imposed before separating events into the ABCD and EM-BDT regions;
these are referred to as high-\dzero leptons in the paragraphs below.
Analysis dependent selections are then applied to high-\dzero leptons to define signal leptons that are used to construct \acp{SR}, which are used to search for signal-like excesses above SM expectations.
Table~\ref{tab:lepton_sel} summarizes the baseline and signal lepton selection requirements, with more details in the following text.

\begin{table}[t]
\small
\begin{center}
\caption{Overview of electron and muon baseline and signal selections.
The signal selection is applied on top of the baseline selection.
The ABCD identification and \pt requirements depend on the SR and on which triggers are used.
$\Delta\pt/\pt = (\pt^{\text{track}} - \pt^e)/\pt^e$ is the relative difference between the track \pt and the calorimeter \et.
$N_{\text{miss}}$ is the number of missing hits.
The $\chi^2$ variables are all reduced by the number of degrees of freedom (n.d.f.), and are for the Inner Detector (ID) track or Combined Muon (CB) track.
$N_{\text{prec}}$ is the number of precision layers, and $N_{\phi}$ is the number of $\phi$ layers.
The TRT barrel selection is only applied for 2023 data.
}
\begin{tabular}{lccc}
\hline
Selection & \multicolumn{2}{c}{Electrons} & Muons \\
\hline
\hline
\multicolumn{4}{c}{Baseline} \\
\hline
\pt (\gev)      & \multicolumn{2}{c}{> 10}  & > 21 \\
|$d_{0}$| (mm)  & \multicolumn{2}{c}{< 300} & 2 < |$d_{0}$| < 300 \\
|$z_{0}$| (mm)  & \multicolumn{2}{c}{< 500} & < 500  \\
|$\eta$|        & \multicolumn{2}{c}{$|\eta|< 1.37 \text{ OR } 1.52 < |\eta| < 2.47$} & < 2.5  \\
Identification  & \multicolumn{2}{c}{LLP VeryLoose} & LLP Medium \\
\hline
\hline
& ABCD Signal (LRT / high-\pt) & EM-BDT Signal & ABCD Signal  (LRT / high-\pt) \\
\hline
Isolation              & Tight\_VarRad &  Tight\_VarRad & PFlow\_Loose\_VarRad \\
Identification         & LLP Loose / -- & LLP Loose  & --\\
\pt (\gev)             & > 31 / > 65   &  --   & -- / > 65\\
$|\dzero|$ (mm)        & > 3         &  > 3  & -- / > 3\\
\deltapt               & $\geq -0.5$ & --    & --\\
$\chi_{\mathrm{ID}}^2/\text{n.d.f.}$   & < 2         & --    & < 2 \\
$N_{\text{miss}}$      & $\leq$ 1    & --    & $\leq$ 1 \\

$N_{\text{prec}}$      & --& --      & $\geq$ 3 \\
$\chi^2_{\text{CB}}/\text{n.d.f.}$   & --& --      & < 3 \\
$N_{\phi}$             & --& --      & > 0 \\
Pass cosmic veto       & --& --      & True \\
Pass TRT barrel (2023) & True & True & True \\
\hline
\end{tabular}
\label{tab:lepton_sel}
\end{center}
\end{table}

Electrons are reconstructed from ID tracks matched to clusters of energy deposited in the EM calorimeter.
If a standard and large-radius track both point to the same EM cluster and lead to the reconstruction of a standard and LRT electron with the
same cluster, the electron satisfying the stricter identification requirement is kept, and in case of a tie, the electron reconstructed with a standard track is kept.
Baseline electrons must have $\pt >10~\gev$, $|\eta|<2.47$, and satisfy a modified version of the \textit{VeryLoose} identification \ac{WP}~\cite{ATLAS:2019qmc} that does not depend on $|\dzero|$, $|\dzero|$ significance, or the number of pixel hits, to avoid suppressing displaced electrons, referred to here as LLP VeryLoose.
Figure~\ref{fig:r22_lepton_effa} shows the efficiency of electrons from a selectron signal model to satisfy the standard and LLP electron identification \ac{WP} as a function of the true lepton $|\dzero|$ value.

High-\dzero electrons, for the purpose of analysis region definition, must additionally satisfy the object overlap removal, have $\pt>31$~\GeV, $3 < |\dzero|<300$~mm, $|z_0|<500$~mm, and fall outside of the \ac{EM} calorimeter barrel-endcap transition region ($1.37<|\eta|<1.52$).
The ABCD signal electron selection additionally requires that the electrons are isolated from nearby tracks and calorimeter energy deposits.
The sum of the \pt of all prompt tracks within a cone, shrinking with \pt (max $\Delta R = 0.2$), and \et\ of all energy depositions within $\Delta R = 0.3$ of the electron must each be less than 6\% of the electron \pt (Tight\_VarRad).
To suppress electrons from random overlaps of hadronic tracks and photon energy depositions, the relative difference between the electron track \pt measurement and that of the electron \pt measured when accounting for the calorimeter energy must satisfy $\deltapt \equiv \dpt \geq -0.5$.
To suppress combinatoric fake tracks, the ID track must have
a $\chi^2$ per degree of freedom (n.d.f.) satisfying $\chi_{\mathrm{ID}}^2/\text{n.d.f.}<2$, where $\chi^2$ is the sum of the squares of the distances between each hit and the track position extrapolated to the given sensor divided by the resolution on this difference, and at most one missing cluster that is included in the track fit (track ``hit'') at a larger radius than the innermost hit.
Signal electrons in regions triggered by the LRT trigger require the stricter LLP Loose identification WP to be consistent with the trigger, with the same modifications as above to the standard \textit{Loose} \ac{WP}~\cite{ATLAS:2019qmc}.
The EM-BDT signal electron selection requires the baseline selection, $3 < |\dzero|<300$~mm, the same %
isolation criterion, and the same LLP Loose identification WP.

Muons are reconstructed from ID tracks matched to muon track segments.
If a standard and LRT track both point to the same muon segment and lead to the reconstruction of a standard and LRT muon
with the same segment, the duplication is resolved by keeping the muon with the smaller difference in $\eta$ between the ID track and the muon track segment extrapolated to the PV.
Baseline muons must have $\pt >21~\gev$, $|\eta|<2.5$, and satisfy a modified version of the \textit{Medium} identification WP~\cite{MUON-2018-03} without requirements on the number of pixel and TRT hits, referred to here as LLP Medium; though, the standard tracks used to reconstruct muons still have a requirement of at least one pixel hit.
Figure~\ref{fig:r22_lepton_effb} shows the efficiency of muons from a smuon signal model to satisfy the standard and LLP muon medium identification \ac{WP} as a function of the true lepton $|\dzero|$ value.

\begin{figure}[t]
\centering
\begin{subfigure}{0.49\textwidth}
\centering
\includegraphics[width=\textwidth]{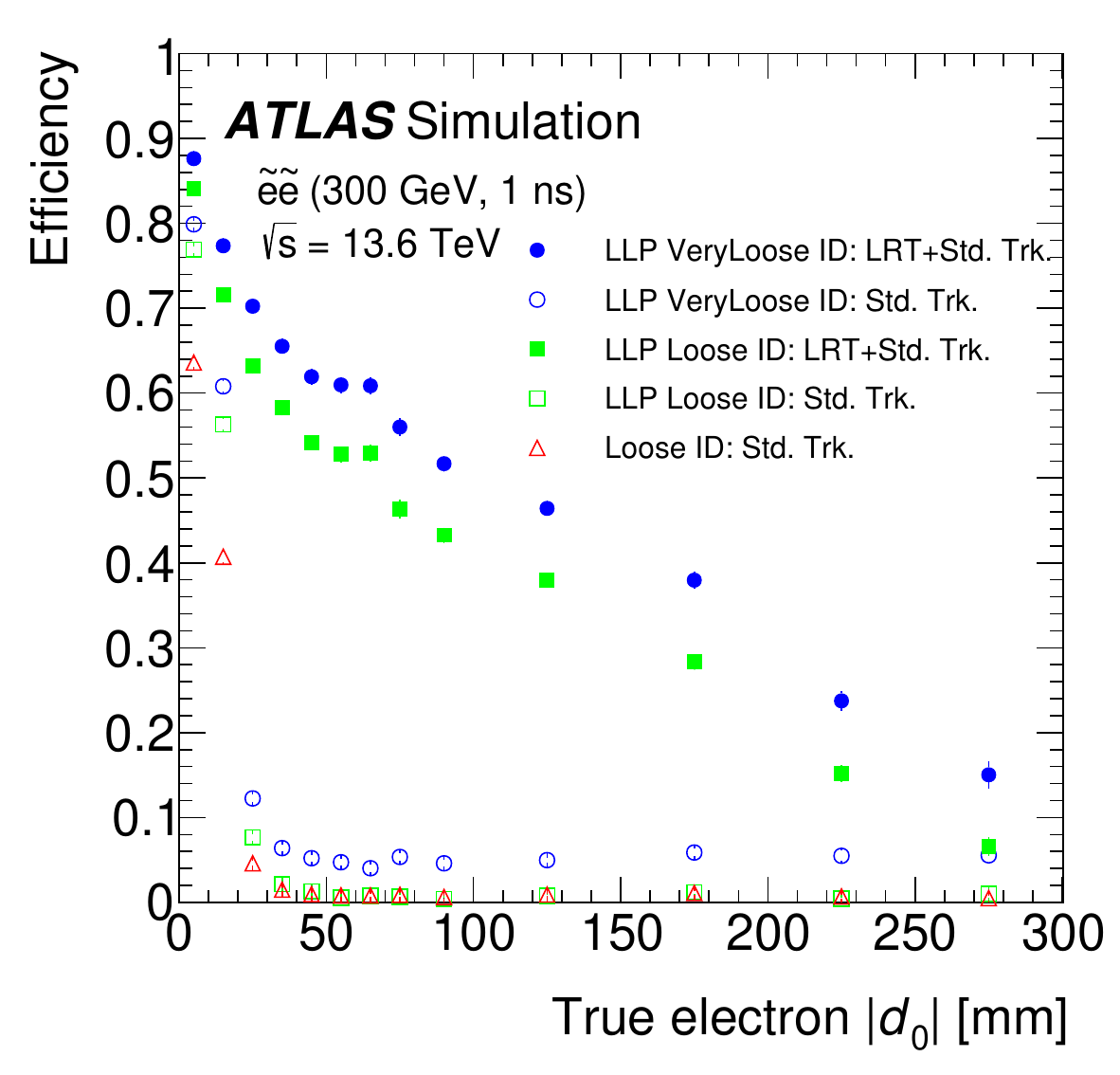}
\caption{}
\label{fig:r22_lepton_effa}
\end{subfigure}
\begin{subfigure}{0.49\textwidth}
\centering
\includegraphics[width=\textwidth]{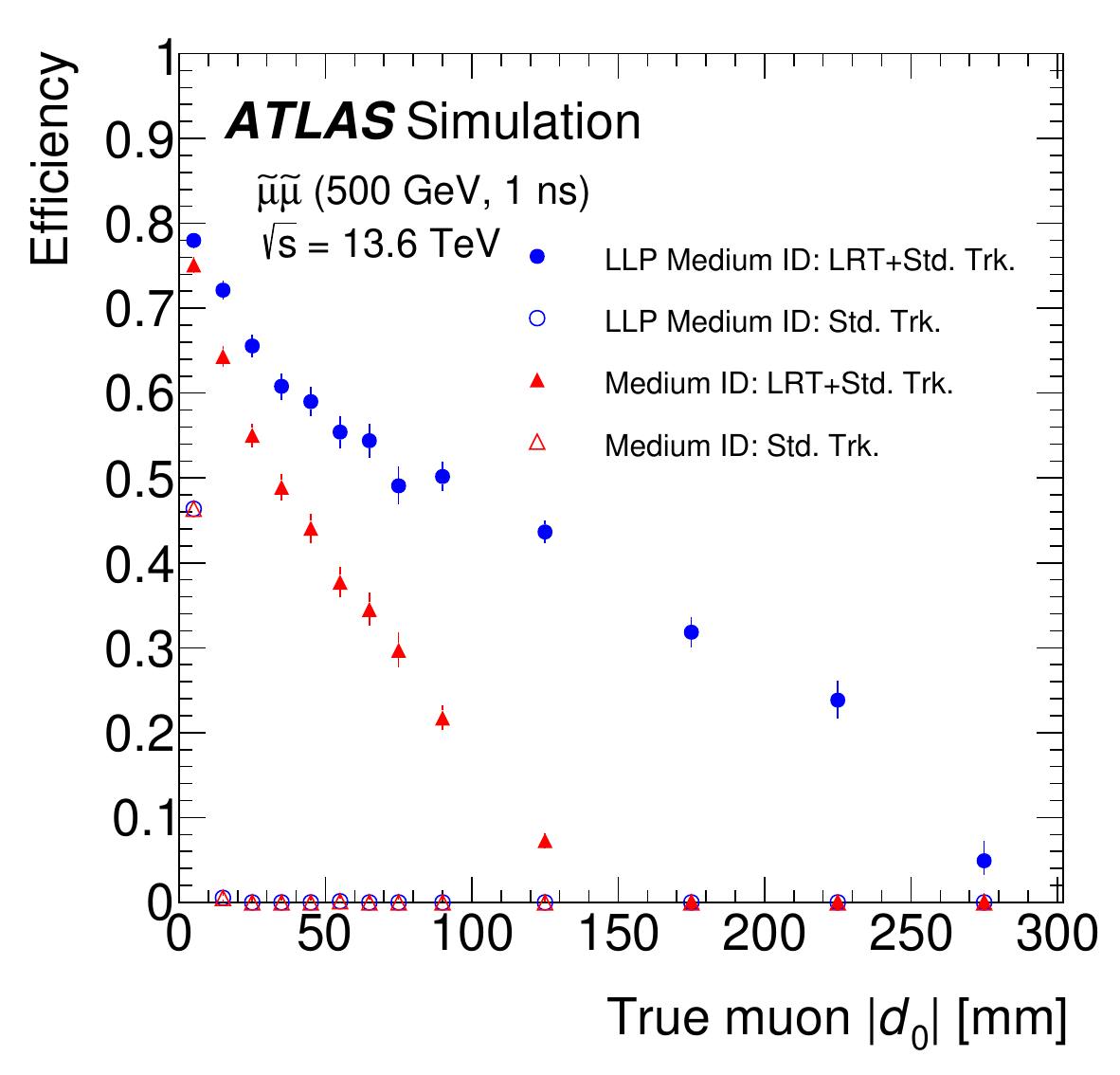}
\caption{}
\label{fig:r22_lepton_effb}
\end{subfigure}
\caption{Efficiency to reconstruct and identify (a) electrons and (b) muons using \acf{LRT} and standard tracking (Std. Trk.) with respect to generator-level leptons in simulated GMSB slepton signal models with a slepton lifetime of 1~ns and a selectron (smuon) mass of 300~(500)~\GeV.
The efficiencies for the various LLP working points are shown
as a function of the true transverse impact parameter $|\dzero|$ of
the daughter lepton.
}
\label{fig:r22_lepton_eff}
\end{figure}

High-\dzero muons must additionally satisfy the object overlap removal, $2 < |\dzero|<300$~mm, and $|z_0|<500$~mm.
Signal muons are required to additionally be isolated from nearby tracks and calorimeter energy
depositions using an algorithm that exploits particle flow to associate tracks with their energy clusters~\cite{MUON-2018-03}.
The sum of the \pt of all prompt tracks within a cone, shrinking in \pt (max $\Delta R = 0.3$), of the muon is added to 40\% of the sum of the \et\ of energy depositions within $\Delta R = 0.2$ that are not matched to tracks and the total is required to be less than 16\% of the muon \pt (PFlow\_Loose\_VarRad).
Several additional requirements are imposed to reject spurious muons from combinatoric fake tracks.
Muons must be measured in at least three MS precision tracking layers and have at least one high-precision $\phi$ measurement.
The ID track must have $\chi^2_{\mathrm{ID}}/\text{n.d.f.}<2$ and at most one missing hit at a larger radius than the innermost hit, and the combined muon (CB) track must satisfy $\chi^2_{\mathrm{CB}}/\text{n.d.f.}<3$.
To reject cosmic-ray muons, a cosmic muon veto is applied that is similar to the one used in the ATLAS Run 2 displaced leptons search~\cite{ATLAS:2020wjh} and originally
developed in the ATLAS Run 2 search for a displaced vertex and a displaced muon~\cite{SUSY-2018-33}.
A muon is tagged as originating from a cosmic-ray muon, and thus vetoed, if there are MS segments on the opposite side of the detector along its trajectory,
within $\eta$ and $\phi$ windows of $|\eta_\mu+\eta_{\mathrm{MS}}| < 0.01$ and $\pi-|\phi_\mu-\phi_{\mathrm{MS}}| < 0.08$,
where the $\mu$ and $\mathrm{MS}$ subscripts refer to the reconstructed muon and MS segments, respectively.
A geometric correction is applied to the $\eta$ value used to take into account tracks that do not pass through the PV.

Photons are reconstructed from EM calorimeter depositions. Baseline photons must have $\et>30~\gev$, fall within the
barrel ($|\eta|<1.37$) or endcap ($1.52<|\eta|<2.5$) regions, and satisfy the \textit{Loose} photon identification WP~\cite{ATLAS:2019qmc}.
Signal photons must also be isolated from nearby tracks and calorimeter depositions~\cite{ATLAS:2019qmc}. %
The sum of the \pt of all prompt tracks (\et\ of all energy depositions) within $\Delta R = 0.2$ of the photon must be less than 5\% (6.5\%) of the photon \et.

Jets are reconstructed with the particle-flow technique~\cite{PERF-2015-09,ref:fastjet} using the $\mathrm{anti-}k_t$ algorithm with a radius parameter of 0.4~\cite{Cacciari:2008gp}. In this analysis, jets are only used for overlap removal and, for the EM-BDT search only, to evaluate the \MET. Baseline jets must satisfy
$\pt>20~\gev$ and $|\eta|<2.8$. To reject jets from pileup interactions, a jet vertex tagger algorithm employing a neural network is applied using a WP with an approximately fixed efficiency as a function of jet \pt~\cite{JETM-2020-03}.

The value of \MET is defined as
the magnitude of the negative vector sum of the transverse momenta of all distinct reconstructed objects as well as tracks not associated with any reconstructed objects \cite{JETM-2020-03}.
When calculating the value of \MET, both prompt and displaced leptons are included.

The analysis requires one of the following three conditions to be true: at least one baseline electron, at least two baseline muons, or at least two baseline photons.
Events must pass one of the triggers described in Section~\ref{sec:datamc},
which depend on the final state and the $\pt$ of the leptons and photons. %
During 2023, part of the TRT barrel was not in operation, in a region defined by $-1.759 < \phi < -1.257$ and $|\eta| < 1.37$.
Events with electrons or muons in this $\eta-\phi$ region are vetoed in 2023 data only.

\begin{figure}[b]
\centering
\includegraphics[width=1\textwidth]{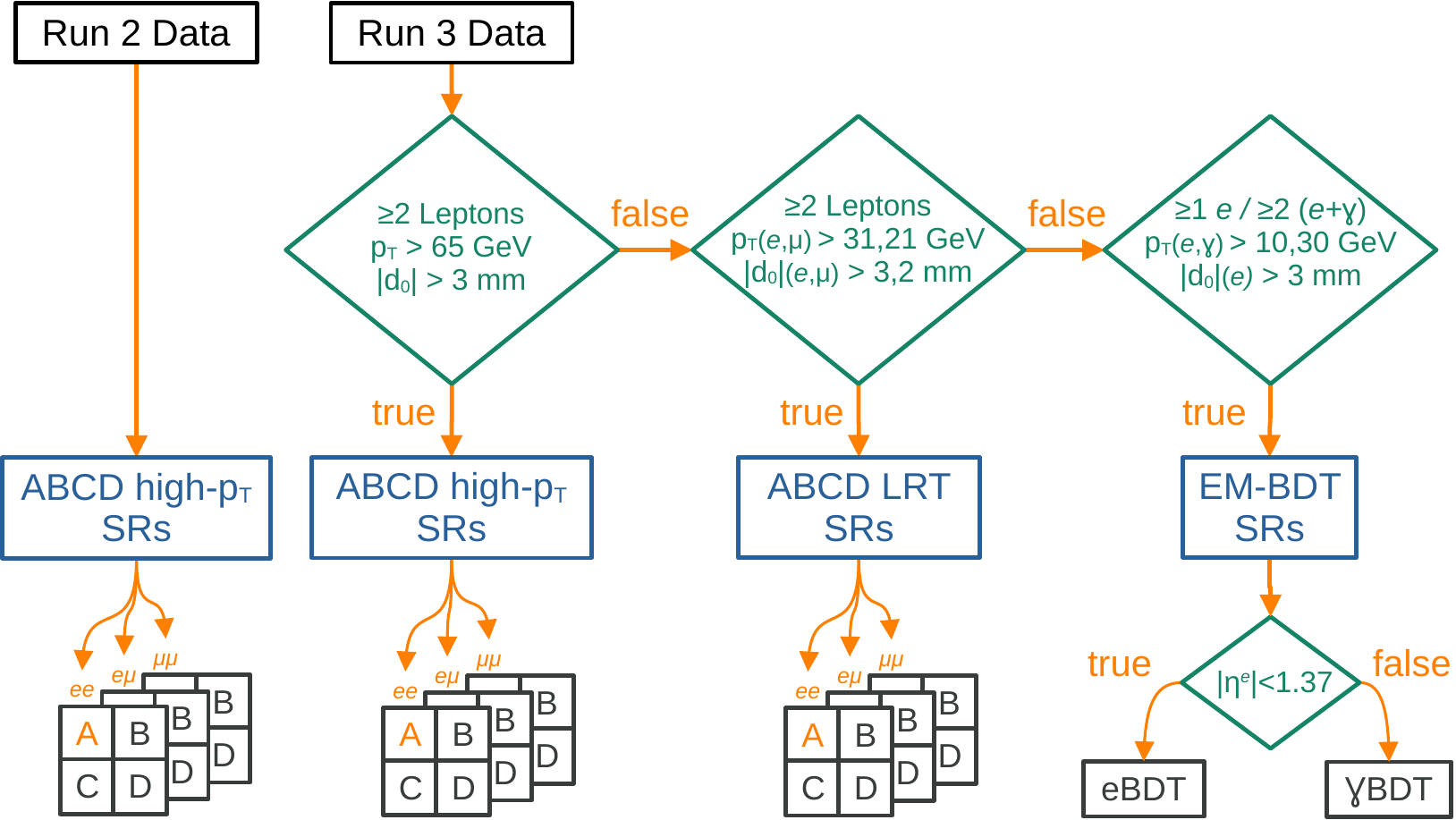}
\caption{Flowchart of the event categorization into the ABCD high-\pt, ABCD LRT, and EM-BDT SRs.
The Run~2 and Run~3 ABCD high-\pt SRs are separate.
}
\label{fig:flowchart}
\end{figure}

Events are categorized into orthogonal
ABCD high-\pt, ABCD LRT, and EM-BDT regions in cascading order based on the number of high-\dzero leptons, as defined above.
If there are at least two such leptons with $\pt>65~\GeV$, events fall into the ABCD high-\pt regions.
Failing that, if events have at least two such leptons with $\pt>31(21)$~\GeV for electrons (muons), events fall into the ABCD LRT regions.
Remaining events fall into the EM-BDT regions.
The flowchart in Figure~\ref{fig:flowchart} illustrates the categorization of events.
Additional selections, defined in Sections~\ref{sec:abcd} and \ref{sec:embdt} for the ABCD and EM-BDT approaches, respectively, are used to define \acp{SR} that are sensitive to BSM signals, \acp{CR} that are used to estimate the backgrounds in the SRs, and \acp{VR} that are used to test the robustness of the background estimation methodology. The SRs, CRs, and VRs for each channel are mutually orthogonal.


%
%
\FloatBarrier
\section{ABCD analysis}
\label{sec:abcd}


The ABCD analysis defines \acp{SR} for three dilepton flavor combinations,
$ee$, $\mu\mu$, and $e\mu$, using the flavors of the two leptons with the highest \pt.
Backgrounds are estimated using two ABCD-based estimations, one for \acrfull{FHF} leptons and one for cosmic-ray muons, which make use of leptons passing and failing certain quality criteria.
These are estimated separately because, while they are uncorrelated with each other, different sets of background suppressing requirements need to be inverted to populate the background estimation regions.
Cosmic-ray muons are only relevant for the $\mu\mu$ final state.
Region A, shared for the FHF and cosmic-ray muon estimates, is the SR where both leptons pass all analysis requirements, regions B and C have one lepton passing the criteria and the other one failing them, and region D contains events where both leptons fail the criteria, which are different for the FHF and cosmic-ray muon estimates.
When the criteria used on each lepton are uncorrelated, the estimate of the background in the SRs reduces to
\begin{equation*}
N_{A}^\text{predicted} = \frac{N_{B}^\text{FHF}\times N_{C}^\text{FHF}}{N_D^\text{FHF}} + \frac{N_{B}^\text{Cosmics}\times N_{C}^\text{Cosmics}}{N_D^\text{Cosmics}},
\end{equation*}
where $N_{B/C/D}^{\mathrm{FHF}}$
and $N_{B/C/D}^{\mathrm{Cosmics}}$ are the event yields in the various control regions
for the FHF and cosmic-ray backgrounds, respectively, for a given lepton flavor combination.
The B, C, and D regions for a given \ac{SR} are included in the fits to determine the backgrounds for model-independent limits and BSM exclusions, as described further in Section~\ref{sec:result}.

\subsection{ABCD event selection}
\label{sec:abcd:sel}

The selection of events for the ABCD analysis begins
with the common preselection requirements described in
Section~\ref{sec:objects}. The ABCD search requires the presence of at least two signal leptons ($e$ or $\mu$) separated
by $\Delta R(\ell_1,\ell_2)>0.2$ and considers two orthogonal sets of SRs.
The high-\pt SRs use a similar selection as the Run~2 search, relying on photon and MS-only muon triggers
and requiring both leptons to have $\pt>65$~\GeV and $|\dzero|>3$~mm.
The Run~3 high-\pt \acp{SR} benefit from the addition of new photon plus MS-only and di-MS-only-muon triggers, listed in Table~\ref{tab:triggers}.
Separate \acp{SR} are defined for $ee$, $\mu\mu$, and $e\mu$ events in Run~2 and Run~3, resulting in six orthogonal ABCD high-\pt SRs.

The ABCD LRT SRs use the novel LRT triggers that were introduced in Run~3 and require electrons (muons) with $\pt>31$ (21)~\gev.
The muon $|\dzero|$ threshold is also reduced to 2~mm.
Electrons are required to pass the stricter LLP Loose WP to be consistent with the LRT trigger.
Events satisfying the high-\pt SR selection are vetoed.
Three SRs exploiting the LRT triggers are defined for $ee$, $\mu\mu$, and $e\mu$ events, in Run~3 only, leading to a total of nine orthogonal SRs considered in the ABCD search.
For the three $\mu\mu$ SRs, an additional cosmic-ray veto
rejects events with back-to-back fully reconstructed muons, $\mu_{1}$ and $\mu_{2}$, with $|\eta(\mu_{1})+\eta(\mu_{2})|<0.02$ and $\pi-|\phi(\mu_{1})-\phi(\mu_{2})|<0.01$, taking advantage of the better angular resolution of fully reconstructed muons at the PV, compared to extrapolated MS segments.

To maintain broad sensitivity to BSM physics scenarios besides the GMSB long-lived slepton and dark sector models
considered in this paper, no additional kinematic selections are applied. Due to the low rates of displaced leptons
in data, these selection criteria are enough to reduce the expected background yields in all SRs to less than one event.

Example total acceptances, summing over the ABCD \acp{SR} lepton flavors, for the stau model with a 100~\GeV mass and 0.01~ns lifetime in the high-\pt and LRT SRs are 0.001 and 0.012, respectively.
For the smuon model with 800~\GeV mass and 0.1~ns lifetime, the total acceptances are 0.19 and 0.06 for the high-\pt and LRT SRs, respectively.

\subsection{ABCD background estimation}
\label{sec:abcd:bkg}

Two sources of background are considered in the ABCD analyses.
The \ac{FHF} lepton background contributes to all three lepton flavor channels and includes combinatoric fake tracks, random overlaps of hadronic tracks with EM clusters or MS tracks, photon conversions, and decays of heavy-flavor (HF) bottom and charm mesons.
The cosmic-ray muon background contributes to $\mu\mu$ events only.
Since these processes are not necessarily well-modeled in simulation, they are estimated from data \acp{CR} using ABCD methods that closely follow the methodology of the Run~2 ATLAS displaced leptons search~\cite{ATLAS:2020wjh}. %
The final background estimates and results for each \ac{SR} are obtained from a simultaneous fit of the four ABCD regions that includes any signal contamination.
The signal contamination is mostly less than 10\% in the B and C regions for the \ac{FHF}, and always less than 1\% in the denominator region D.
It is only larger in a few B and C regions depending on the signal model.
Thus any contamination would lead to a lightly overestimated background and a weaker analysis sensitivity.
For the cosmic-ray muon estimation, the signal contamination is negligible in each region.

The background estimate for FHF leptons uses \acp{CR} in which some of the quality criteria for one lepton, the other lepton, or both, are inverted.
A failing electron is one that fails at least one of the requirements from Table~\ref{tab:lepton_sel} on \dpt, ID track $\chi_{\mathrm{ID}}^2/\text{n.d.f.}$, and the number of missing track hits, while satisfying all other ABCD signal electron requirements of Section~\ref{sec:objects}.
A failing muon is one that fails at least one of the requirements from Table~\ref{tab:lepton_sel} on the number of precision tracking layers, the presence of a high-precision $\phi$ measurement,
ID track $\chi_{\mathrm{ID}}^2/\text{n.d.f.}$,
combined muon $\chi_{\mathrm{CB}}^2/\text{n.d.f.}$, and the number of missing track hits, while satisfying all other ABCD signal muon requirements of Section~\ref{sec:objects}.
In the $\mu\mu$ high-\pt regions only, the observed yields in the ABCD \acp{CR} are too small to provide individual predictions for Run~2 and Run~3.
Therefore the yields in the ABCD regions for these two datasets are combined and the predicted FHF background is scaled by the
fractions of the total luminosity accounted for by each dataset, namely 0.71 for Run~2 and 0.29 for Run~3.
The predicted \ac{SM} background yields in each of the nine ABCD SRs is less than 0.2 events.
As an example, for the high-\pt Run~3 $ee$ SR, the FHF B, C, and D region yields are 3, 6, and 278 respectively.
This results in a predicted background of $3\cdot6 / 278 = 0.067^{+0.092}_{-0.041}$ assuming no signal contamination and ignoring the SR yields, which are included as a constraint in the fit for the final results.

This method assumes that the quality criteria of the two FHF leptons are uncorrelated with each other.
The correlation coefficient, as measured in \acp{VR}, varies between -0.05 and -0.1.
Two sets of \acp{VR} are defined to account for any systematic biases of the background estimation methodology, including potential correlations between the quality criteria of the leptons.
The VR-fake regions are enriched in fake-lepton backgrounds by inverting the requirement on \dpt\ for electrons
and the CB track $\chi_{\mathrm{CB}}^2/\text{n.d.f.}$ requirement for muons.
This makes these VRs a subset of region D, but orthogonal to regions A, B, and C.
The VR-HF regions are enriched in heavy-flavor leptons by requiring at least one lepton to fail the isolation criteria.
Some requirements are removed in order to increase the number of events in the heavy-flavor VRs: \dpt\ in the $ee$ and $e\mu$ channels; $\chi^2_{\text{CB}}/\text{n.d.f.}$ in the $\mu\mu$ channel; and $N_{\text{prec}}$ and $N_{\phi}$  in the $e\mu$ channel.
The observed yields in the nine VR-fake and VR-HF regions corresponding to the nine \acp{SR} are consistent with the background expectations obtained by performing the ABCD estimate in these \acp{VR}.

For cosmic-ray muons, the ABCD \acp{CR} are defined by inverting the MS quality requirements (number of precision tracking layers $N_{\text{prec}}$, number of high-precision $\phi$ measurements $N_{\phi}$, or combined muon $\chi_{\mathrm{CB}}^2/\text{n.d.f.}$) or by inverting the cosmic veto relying on the presence of MS segments on the opposite side of the detector.
The correlation coefficient between these quantities is -0.03, computed in events with two baseline muons.
These are inverted for the muon in the top half of the detector, which is more likely to be mis-reconstructed than the muon in the bottom half since it travels in the opposite direction of particles that originate from the PV.
For example, region B has a cosmic tagged top muon and region C has a top muon failing at least one MS quality requirements.
Applying the additional cosmic-ray veto on back-to-back fully reconstructed muons eliminates almost all of the events in these regions.
Therefore the cosmic-ray muon background is estimated without this additional cosmic-ray veto applied and then scaled by the expected efficiency $T=(3.7\pm2.6)\times 10^{-3}$ for muons to survive the veto, evaluated using the events that fall in the superset of regions B, C, and D.
The expected background yields are of the order $10^{-3}$ in the three $\mu\mu$ SRs.

This background method is validated in VRs enriched in cosmic-ray muons, in which the cosmic veto relying on the presence of MS segments on the opposite side of the detector as the bottom muon is replaced by the requirement that this muon has MS segments satisfying $|\eta_\mu+ \eta_\text{MS}| < 0.01 $ and $\pi-|\phi_\mu - \phi_\text{MS}| < 0.08$ but no such segments satisfying  $|\eta_\mu+ \eta_\text{MS}| < 0.002$ and $\pi - |\phi_\mu - \phi_\text{MS}| < 0.02$.
To enrich the VR in cosmic-ray muons, the additional cosmic-ray veto based on the two fully reconstructed muons is not applied.
The observed yields are consistent with the background expectations.

\subsection{ABCD systematic uncertainties}
\label{sec:abcd:syst}

While the sensitivity of the ABCD  analysis is dominated by statistical uncertainties, systematic uncertainty are derived based on the agreement, or closure, of the VRs.
Since these VRs, like the SRs, have limited numbers of events, the larger of the non-closure and the statistical uncertainty is used as the systematic uncertainty.
Additional uncertainties common with the EM-BDT analysis are described in Section~\ref{sec:syst}.

The systematic uncertainty in the FHF lepton background is based on the agreement between observed yields versus expected backgrounds in the VR-fake and VR-HF regions, testing the closure of the method in regions orthogonal to the SRs.
For each region, the larger of two quantities is taken as the systematic uncertainty: the deviation from one of the ratio of the observed yield to the expected background, or the statistical uncertainty in the expected background. The values are extracted from VR-fake and VR-HF separately and then added in quadrature.
Due to the small rate of fake muons, the $e\mu$ and $\mumu$ VRs all had less than five observed events.
In all cases, the statistical uncertainty is equal to or larger than the non-closure.
This results in positive uncertainties of $50/190/150$\% ($ee$/$e\mu$/$\mu\mu$) for the Run~2 high-\pt regions; $40/160/150$\% for the Run~3 high-\pt regions; and $130/150/240$\% for the Run~3 LRT regions.

Two sources of systematic uncertainty in the cosmic-ray muon background are considered. The first source is due to the assumption that the muon in the top half of the detector is the one that is mismeasured.
The background is estimated by instead inverting the cosmic veto and/or MS quality variables ($N_{\text{prec}}$, $N_{\phi}$, and $\chi^2_{\text{CB}}/\text{n.d.f.}$) of the bottom muon and taking the relative difference with respect to the nominal result, resulting in uncertainties of 28\%, 2\%, and 41\% in the Run~2 high-\pt, Run~2 high-\pt, and Run~3 LRT SRs respectively.
The cosmic-ray background is also estimated by inverting two of the three MS quality variables and taking the relative difference with respect to the nominal estimate.
The largest deviation of the three pairs is used, resulting in uncertainties of 20\%, 38\%, and 82\% in the Run~2 high-\pt, Run~2 high-\pt, and Run~3 LRT SRs respectively.


%
\FloatBarrier
\section{EM-BDT analysis}
\label{sec:embdt}


The EM-BDT analysis seeks to discriminate between
displaced electrons and the background, which is
dominated by prompt particles. In the interpretation
of the results for
the GMSB long-lived slepton model used as a benchmark
(see the diagram in Figure~\ref{fig:signal_diagrams}),
the EM-BDT analysis is
applied to the selectron and stau channels, since
either can produce displaced electrons in the final
state.

As an electron arising from the decay of a long-lived
parent is both displaced and delayed, there exist
significant correlations between many
relevant observables, motivating the introduction of machine learning techniques in the development of a
multivariate discriminant.
An electron from an LLP decay that occurs too far into the detector may not have an identifiable track. In such a case, the electron is reconstructed as a photon, provided the decay is before the calorimeter.
Two different BDTs
are employed, each using variables that are verified with comparisons of $Z\rightarrow ee$ data and MC simulation to be reasonably well modeled.
The ``\eBDT''
seeks to identify
displaced electrons from among those particles
reconstructed as electrons, while the ``\gBDT''
seeks to identify
displaced electrons from among those particles
initially
reconstructed as photons. The \gBDT\ extends the
sensitivity of the analysis to events where the decay
producing the electron occurs in
front of the EM calorimeter but
farther from the beamline than the
300~mm limit where LRT is applied; in such
cases, it is still possible to identify the
displaced electron from the EM calorimeter information
only.

The discriminating variables for the \eBDT\ include
tracking information, EM calorimeter variables,
and combinations of the ID and LAr measurements.
The tracking variables include the track fit quality
$\chi^2$ value, the number of pixel hits on the track,
the number of missing layers in the track,
and the transverse and longitudinal impact
parameters with respect to the PV position.  The EM
calorimeter variables include the LAr pointing measurement ($z_\text{EM}$), the absolute value of the LAr timing measurement,
and two LAr shower shape variables, namely the fraction of the EM shower energy
in the first
and third longitudinal layers ($f_1$ and $f_3$, respectively).
The combined variables include
\dpt
and the difference between $z_0$ and
$z_\text{EM}$; both of these combined
variables are intended to help reject
backgrounds from fake electrons.
The background
rejection of the \eBDT\ is enough that the
analysis can be performed with only one
reconstructed electron in the event, thereby greatly
extending the analysis sensitivity to longer slepton
lifetime values, where only one selectron (or stau) has
decayed before reaching the EM calorimeter. In events
with more than one reconstructed electron, or with
one electron and one or more photons, the \eBDT\
is applied to the leading electron. Thus, the \eBDT\
is an ``object-level'' discriminant, using only
measurements of the electron itself and not of other
properties of the event, and thereby functions as a
``displaced electron tagger.'' This feature of
the \eBDT\ makes it rather model-independent.

The \gBDT\ searches for displaced electrons reconstructed as photons. The lack of tracking information and higher background rates for photons than electrons makes a single photon region infeasible.
Instead, the \gBDT\ requires at
least one additional electron or photon.
The variables used are the
pointing and timing measurements of the two
objects, as well as the differences between them.
In addition, the value of \MET\ in the event is used.
This choice is motivated by the gravitinos produced in
the baseline slepton model in order to extend the reach of
the analysis to longer slepton lifetime values, but does make the \gBDT\ more
model-dependent than the \eBDT.

\subsection{EM-BDT event selection}
\label{sec:embdt:sel}

The selection of events for the EM-BDT analysis begins
with the common preselection requirements described in
Section~\ref{sec:objects}.
The EM-BDT event selection is
made orthogonal to that of
the ABCD analysis regions by vetoing dielectron
events passing the ABCD \pt\ and $|\dzero|$\ requirements.
The selected EM-BDT events are divided into several distinct final states,
according to the multiplicity of reconstructed EM objects (electrons or photons)
in the event. Each final state is selected by a set of triggers, as described
in Table~\ref{tab:triggers}.

The \eBDT\ is used to analyze events with at
least one electron in the barrel calorimeter $(|\eta| < 1.37)$. These
include single-electron events, but also events with an electron in
the barrel calorimeter plus either additional electrons or photons.
In events
with more than one reconstructed electron, the leading electron in the
barrel is selected as the input object to the \eBDT\ algorithm.
The electron is required to have \maxEcell $> 5$~\GeV to reduce the impact
of noise on the calorimeter
timing, and an absolute value of its calorimeter time less than 12.5~ns,
to reject backgrounds from other bunch crossings.

Given the large backgrounds in single-photon event samples
as well as the lack of tracking information for photons,
the \gBDT\ is used to analyze only events with at least two reconstructed
EM objects; the first must be a photon passing the baseline requirement,
while the second can be another photon or an electron which fails
the \eBDT\ requirements.
Electrons are treated as photons in this case, and do not have the transverse impact parameter requirement in Table~\ref{tab:lepton_sel}.
At least one EM object must be in the barrel calorimeter $(|\eta| < 1.37)$.
Both EM objects are required to have \maxEcell $> 5$~\GeV and an absolute value of its calorimeter time less than 12.5~ns.
A requirement of $|z| < 2000$~mm for both EM
objects eliminates beam-induced backgrounds
that can fake the photon signature.

For a selectron with a mass of 200~\GeV and a lifetime of 30~ns, total acceptances of the EM-BDT event selection are 0.0022 for the \gBDT\ and 0.01 for the \eBDT.
The sum of these are two orders of magnitude larger than the sum of the ABCD \ac{SR} acceptances.

\subsection{EM-BDT background estimation}
\label{sec:embdt:bkg}

The background in the EM-BDT analysis is dominated by
prompt particles that are mismeasured or
poorly reconstructed. Such effects cannot be sufficiently
precisely simulated, and therefore the background
estimation is necessarily
entirely data-driven. The estimate is
generated by dividing the data into three orthogonal
regions: a \ac{CR}, which can be used estimate the background normalization and
BDT distribution shape, a \ac{VR}, which can be used to
validate the background estimate before unblinding the \ac{SR}
data, and the SR, in which the search will ultimately be performed.

The EM-BDT analysis exploits the fact that, for prompt
background, the distribution of the arrival time at the
EM calorimeter should, due to resolution effects,
be symmetric about zero, as seen in Figure~\ref{fig:timing_d0}.
In contrast, signal events
should be delayed, and therefore have primarily positive
arrival times. As a result, the CR can be defined
for the \eBDT\ (\gBDT) by events where the leading electron has (both photons have)
a negative time, while the SR is defined by events where the
leading electron has (both photons have) a positive time.  The VR is defined
by kinematic selections that aim to achieve a dominance
of background over signal by selecting
events with enhanced SM $W$ or $Z$ boson production; the
VR includes single-electron events with a
transverse mass of the lepton and \met system in the range from 60--100~\GeV, and
dielectron events with a dielectron mass in the range
from 80--100~\GeV.
To guarantee orthogonality among the three regions, the CR and SR definitions include vetoes of any events that fall within the VR selection.

For both the \eBDT\ and the \gBDT, the BDT is trained
to separate the signal, as modeled with MC simulation, from
the background, as modeled by the CR data.
Signal models of various masses and lifetimes $\geq1$~ns were used for training in order to maximize the sensitivity to longer lifetimes.
To make use of the entire sample of the MC simulation and the CR data, the signal sample and data used for training are split randomly into three ``folds'' and training is performed on each group individually. It is then required that the score assigned to a particular event during evaluation is determined using a fold other than the one assigned to that event in training.
The final EM-BDT results for various slepton masses and lifetimes are obtained by signal-plus-background fits,
including associated statistical and systematic
uncertainties, of the unblinded SR data to the weighted
sum of the CR-derived background template and signal template shape from MC simulation.
The signal contamination in the CR is found to be less than one percent of the yield.
Still, to account for signal contamination in the CR, particularly at low lifetimes where the signal has a similar shape in the CR and SR, the signal template in the fit is set to be the difference of the positive and negative timing BDT score distributions, i.e.\ the difference between the SR and CR templates.
Before unblinding the SR data, the
BDT distributions were validated by performing fits
to the background-dominated VR.

The BDTs are defined such that their output values range from $-1$ to $+1$, with signal-like (background-like) events skewed toward positive (negative) BDT scores.
The final statistical analysis uses a simultaneous fit to binned~\eBDT\ and \gBDT\ distributions, where the normalization is allowed to float.
Before unblinding, the binning used for the BDT scores was subject
to an optimization procedure that sought to maximize the signal sensitivity.
To ensure a non-zero background estimate could be made for each bin, the choice of bins was subject to the constraint that there exist at least one CR event in each bin.
The \eBDT\ and \gBDT\ binnings were optimized separately.
However, for both BDTs the very background-like events with BDT scores less than zero are not used in the fit,
and the events with positive BDT scores are divided into five non-overlapping bins.

\subsection{EM-BDT systematic uncertainties}
\label{sec:embdt:syst}

While the sensitivity of the EM-BDT analysis
is dominated by statistical
uncertainties, several sources of systematic uncertainty
are considered.
The background normalization is determined in the fit, and therefore no corresponding systematic
uncertainty is needed. However, systematic uncertainties
that are considered can impact
the results of the final fit by either
impacting the shapes of the BDT score distribution of the signal
or the background, or by affecting only the signal normalization.
Additional uncertainties common with the ABCD analysis are described in Section~\ref{sec:syst}.

The largest systematic uncertainty in the shape of the
background BDT distribution arises due to the assumption that
the CR data, which has negative time values, describes accurately
the expected background in the SR region, which has positive times.
The systematic uncertainty in this assumption is determined by
using the VR data, where the signal contribution is negligible and
therefore the symmetry of the background around $t = 0$~ns can
be tested quantitatively. Comparing the BDT distribution of the VR data
with negative times with that of the VR data with positive times,
the relative discrepancies increase monotonically toward the
higher BDT score bins, and reach as high as 33\% (84\%)
in the highest bin of the \eBDT\ (\gBDT) score.  These
``non-closure'' uncertainties are symmetrized about the
central prediction and included as systematic
uncertainties in the BDT background shape.

A systematic uncertainty in the shape of the signal BDT distribution arises due to uncertainties in the
smearing of the time resolution.
The MC simulation does not accurately describe the timing resolution in data, and
therefore the signal MC events are subjected to an additional smearing of the time.
The smearing procedure consists of taking the raw time from the MC simulation, which already contains a model of the observed noise in the calorimeter, and adding additional correlated and uncorrelated components.
The correlated component corresponds to the beam spread, and is therefore shared by all objects in the same event.
The uncorrelated component is applied separately to each object and corresponds to different electronic and pileup noise assumptions and other global effects observed in comparisons between data and MC simulation.
To determine an uncertainty in the signal BDT shape, the smearing is varied up and down within the experimental uncertainties, and the signal BDT shape is re-evaluated.

%

%


%
\FloatBarrier
\section{Common systematic uncertainties}
\label{sec:syst}

Several sources of systematic uncertainties are evaluated for the prediction of signal model yields, both from the theoretical and experimental sides.
The factorization and renormalization scales, $\alpha_S$, parton distribution functions, and QCD radiation models are all varied to assess the impact on the simulated acceptance in the \acp{SR}.
For the GMSB slepton model in the EM-BDT \ac{SR} a 12\% theory acceptance uncertainty is applied on the yield, while in the ABCD \acp{SR} the uncertainty varies between 10--30\% in the excluded region.
For the dark sector model, theory acceptance uncertainties in the yield in the ABCD \acp{SR} also vary between 10--30\% in the excluded regions.
The theory uncertainty in the total predicted cross-section is not included as an additional uncertainty in the fit so that they are separated from experimental uncertainties and may easily be updating with future improvements to the predictions.

Experimental-based uncertainties in the modeling of leptons are determined in two parts.
Electrons and muons are reconstructed with a track and calorimeter or MS information, respectively.
These sub-detectors are sufficiently far from the \ac{PV} relative to the displacements targeted in this search that the sub-detectors performance, and simulation of said performance, should not differ by much from that for prompt leptons.
Thus uncertainties in the tracking and calorimeter or MS information are separated in order to determine each in the most robust way.
Additional uncertainties for displaced leptons are added to those derived for prompt leptons, which cover the calorimeter and MS portion of the leptons.
Standard uncertainties in the modeling of photons are included; there are no additional displacement based uncertainties applied for photons other than that for the smearing of the time resolution described in Section~\ref{sec:embdt:syst}.

Uncertainties in reconstructed objects are propagated to the reconstruction of \MET.
Additional uncertainties in \MET\ come from varying the scale and resolution of the soft term, parallel and perpendicular to the hard term~\cite{JETM-2020-03}.
As \MET\ is used in selecting events with enhanced SM $W$ production to define the EM-BDT VR, a systematic is assessed on the impact of \MET-reconstruction uncertainties in the fitted EM-BDT score distributions.
The migration of events from EM-BDT SRs to the EM-BDT VR was studied after redefining the VR using variations in \MET.
Based on this a 2\% uncertainty in the signal yields in the EM-BDT SRs was assigned.
For the \gBDT, which uses \MET\ as an input variable, an additional uncertainty is evaluated  by propagating the \MET systematic uncertainties into the \gBDT\ score calculation and re-evaluating the \gBDT\ shape.
Uncertainties in \MET\ have no impact in the ABCD analysis since it is not used in the event selection.

The uncertainty in the modeling of \ac{LRT} in offline reconstruction was determined from comparisons of data and \ac{MC} simulation using $K_S^0$ vertices in Ref.~\cite{IDTR-2021-03}.
These are parameterized in \pt\ and radius of the first hit on the track, and vary from 0.3--16.8\% per large-radius track.

Uncertainties in the simulated acceptance of the trigger for standard prompt leptons and photons are considered.  Additional uncertainties in the \ac{LRT}-based trigger acceptance are computed by comparing the trigger tracking efficiency with respect to offline tracks in data and \ac{MC} simulation, as done in Ref.~\cite{TRIG-2022-01}.
This amounts to an additional uncertainty of 11\% for electrons and 2\% for muons on top of the standard trigger uncertainties.

Standard corrections to the modeling of reconstruction, identification, and isolation requirements for leptons were computed with prompt $Z\rightarrow\ell\ell$ events using the same methods as described for electrons in Ref.~\cite{ATLAS:2019qmc} and muons in Ref.~\cite{MUON-2018-03}.
Uncertainties in the corrections for electrons are generally less than 11\% for $\pt>30~\GeV$; they rise to 70\% for electrons with $\pt<20~\GeV$.
For muons the uncertainties in the corrections are generally less than 3\%.
An additional uncertainty in electron identification as a function of $|\dzero|$ is assessed by comparing the relative loss in efficiency between the VeryLoose and Loose WPs of signal \ac{MC} simulation; this amounts to an uncertainty of 16\% per electron.
For muons, the MS is sufficiently far away from the \ac{PV}, and their identification does not make use of a likelihood discriminant.
Thus, no additional displacement-based uncertainty is applied.

Part of the uncertainties in the ABCD background predictions come from the propagation of statistical uncertainties in the yield in each region.
These range from plus 0.003--0.4 events on a predicted central value of less than one event in each SR, Table~\ref{tab:UL}.
The negative variation is capped so that the expected yield is not negative.
The other set of uncertainties are those described in Section~\ref{sec:abcd:syst} on the \ac{FHF} and cosmics muon background predictions.
The uncertainty on the \ac{FHF} prediction ranges from 30--210\%.
The uncertainties on the cosmics muon background prediction, 2--81\%, have no impact on the analysis as the prediction is of the order $10^{-3}$.
The results in the ABCD SRs are dominated by Poisson statistical uncertainty on the yield.
The dominant uncertainty in the signal prediction in the ABCD \acp{SR} depends on the region and signal model.
Typical ranges for uncertainties are: 11--40\% for the number of  \ac{MC} simulation events, 25\% for the signal theory acceptance, 10--40\% for the uncertainty in \ac{LRT}, 1--15\% for muon modeling, and 20--30\% for electron modeling.

The uncertainty in the BDT backgrounds are dominated by the shape uncertainties, except for the most sensitive bin where the number of CR events is low and the statistical uncertainty is large.
The dominant uncertainties for the $e$BDT are the electron modeling at 21\%, the theory acceptance uncertainty at 12\%, and the statistical uncertainty in the number of simulated events in the most sensitive bins at 7\%.
The dominant uncertainties for the $\gamma$BDT are the signal acceptance uncertainty of 12\%, and the statistical uncertainty in the number of simulated events in the two most sensitive bins at 11 and 4\%.
For both the ABCD and EM-BDT regions, the uncertainty in the lifetime reweighting, discussed in the next section, may be the dominant uncertainty when used.

%


\FloatBarrier
\section{Results}
\label{sec:result}


Each of the orthogonal ABCD signal regions is used individually to compare the observed data yield with the expected background, the results of which are used as ``discovery regions'' to set model-independent limits on the production of displaced leptons; for the ABCD high-\pT analyses, these are quoted separately for the Run~2 and Run~3 data, due to the different center-of-mass energies, resulting in a total of nine
orthogonal ABCD discovery regions.
An EM-BDT discovery region is defined as the most signal-like bin of the \eBDT\ distribution (eBDT score $>0.46$),
and is used to set model-independent limits on the production of displaced electrons.
All model-independent limits are derived using pseudo-experiments.
Model-dependent limits on the long-lived slepton model are obtained by performing a simultaneous fit to the \eBDT\ and \gBDT\ distributions, as well as the nine ABCD SRs.
Only the ABCD \acp{SR} are used for the dark sector model.
The EM-BDT \acp{SR} would be sensitive to some semi-hadronic decays, but the MC simulation only includes decays to two leptons; thus it was not included for limit setting.

\begin{table}[t]
\centering
\caption{Expected and observed yields in the signal regions and model-independent upper limits derived using pseudo-experiments.
Left to right: total expected background after a background-only fit, with the combined statistical and systematic uncertainties; number of events observed in data; 95\% CL upper limits on the visible cross-section ($\langle A\epsilon\sigma\rangle_{\text{obs}}^{95}$) and on the number of signal events ($S_{\text{obs}}^{95}$ ).
The sixth column ($S_{\text{exp}}^{95}$) shows the expected 95\% CL upper limit on the number of signal events, given the expected number (and $\pm 1\sigma$ excursions of the expectation) of background events.
The last two columns indicate the confidence level of the background-only hypothesis (CL$_\text{b}$) and discovery $p$-value with the corresponding Gaussian significance ($Z(s=0)$).
CL$_\text{b}$ provides a measure of compatibility of the observed data with the signal strength hypothesis at the 95\% CL limit relative to fluctuations of the background, and $p(s=0)$ measures compatibility of the observed data with the background-only hypothesis relative to fluctuations of the background.
The $p$-value is capped at 0.5.}
\setlength{\tabcolsep}{0.0pc}
\begin{tabular*}{\textwidth}{@{\extracolsep{\fill}}lP{-1}P{-1}cP{-1}P{-1}cc}
\noalign{\smallskip}\hline\noalign{\smallskip}{ Signal Region}  & \multicolumn{1}{c}{Total Bkg.} & \multicolumn{1}{l}{$N_{\mathrm{obs}}$}  & $\langle A\epsilon{ \sigma}\rangle_\text{obs}^{95}$ [fb]  &  \multicolumn{1}{c}{$S_\text{obs}^{95}$}  & \multicolumn{1}{c}{$S_\text{exp}^{95}$} & CL$_\text{b}$ & $p(s=0)$ ($Z$)  \\
\noalign{\smallskip}\hline\noalign{\smallskip}
\noalign{\smallskip}\hline\noalign{\smallskip}
$ee$ high-\pt (Run 2)      & 0.031,0.031     & 0 & $0.02$ &  $3.0$ & 3.0\rlap{$^{ +0.0 }_{ -0.0 }$} & $0.46$ & $ 0.5$~$(0)$ \\[.5ex]%
$ee$ high-\pt (Run 3)      & 0.06,0.05       & 0 & $0.05$ &  $3.0$ & 3.0\rlap{$^{ +0.0 }_{ -0.1 }$} & $0.45$ & $ 0.5$~$(0)$ \\[.5ex]%
$ee$ LRT                & 0.0016\rlap{$^{ +0.0029 }_{ -0.0016 }$}       & 1 & $0.07$ &  $4.1$ & 3.0\rlap{$^{ +0.0 }_{ -0.1 }$} & $0.97$ & $ 0.01$~$(2.2)$ \\[.5ex]%
$\mu\mu$ high-\pt (Run 2)  & 0.02\rlap{$^{ +0.22 }_{ -0.02 }$}       & 0 & $0.02$ &  $3.0$ & 3.0\rlap{$^{ +0.0 }_{ -0.0 }$} & $0.49$ & $ 0.5$~$(0)$ \\[.5ex]%
$\mu\mu$ high-\pt (Run 3)  & 0.01\rlap{$^{ +0.11 }_{ -0.01 }$}       & 0 & $0.05$ &  $3.0$ & 3.0\rlap{$^{ +0.0 }_{ -0.0 }$} & $0.48$ & $ 0.5$~$(0)$ \\[.5ex]%
$\mu\mu$ LRT            & 0.02\rlap{$^{ +0.04 }_{ -0.02 }$}       & 0 & $0.05$ &  $3.0$ & 3.0\rlap{$^{ +0.0 }_{ -0.0 }$} & $0.49$ & $ 0.5$~$(0)$ \\[.5ex]%
$e\mu$ high-\pt (Run 2)    & 0.0016\rlap{$^{ +0.0033 }_{ -0.0016 }$}       & 0 & $0.02$ &  $3.0$ & 3.0\rlap{$^{ +0.0 }_{ -0.1 }$} & $0.50$ & $ 0.5$~$(0)$ \\[.5ex]%
$e\mu$ high-\pt (Run 3)   & 0.004\rlap{$^{ +0.010 }_{ -0.004 }$}       & 0 & $0.05$ &  $3.0$ & 3.0\rlap{$^{ +0.0 }_{ -0.0 }$} & $0.50$ & $ 0.5$~$(0)$ \\[.5ex]%
$e\mu$ LRT              & 0.2\rlap{$^{ +0.4 }_{ -0.2 }$}       & 0 & $0.05$ &  $3.0$ & 3.0\rlap{$^{ +0.1 }_{ -0.1 }$} & $0.45$ & $ 0.5$~$(0)$ \\[.5ex]%
\noalign{\smallskip}\hline\noalign{\smallskip}
EM-BDT        & 1.0\rlap{$^{ +2.3 }_{ -0.9 }$} & $3$ & $0.13$ &  $7.1$ & 5.7\rlap{$ ^{ +1.9 }_{ -0.8 }$} & $0.77$ & $ 0.2$~$(0.9)$ \\[.5ex]%
\noalign{\smallskip}\hline\noalign{\smallskip}
\end{tabular*}
\label{tab:UL}
\end{table}

Only one event is observed among the
nine orthogonal ABCD signal regions, namely
in the SR$ee$-LRT region.
The observed and expected yields in the orthogonal ABCD
signal regions
are presented in Table~\ref{tab:UL}. %
Corresponding model-independent upper limits on the visible BSM cross-sections, the observed and expected limits on the number of non-SM events, and the $p$-values and corresponding
local $Z$ scores are also shown.
The one event observed in
the SR$ee$-LRT region, compared to the
background expectation of
$0.0016^{ +0.0029 }_{ -0.0016 }$  events corresponds to a local significance
of $2.2\sigma$.

The observed BDT distributions in the SRs %
of the
\eBDT\ and \gBDT\ analyses are shown in Figure~\ref{fig:bdt_score}.
Shown superimposed are the background shapes as determined using
the CR data, and example GMSB signal models with selectrons.
The lower panels in the figures show the bin-by-bin ratios of the data divided
by the background, and illustrate that the observed SR data
are consistent
with the background expectation. Defining the EM-BDT discovery region as the most signal-like bin of the \eBDT \ yields a background
expectation of one event. The uncertainty in this background estimate is dominated by the statistical uncertainty resulting from the fact that it is determined from a single CR event in this bin of the \eBDT, plus a systematic uncertainty of $\pm 0.33$ from the other sources of uncertainty described previously.
A total of 3 events are observed in the EM-BDT discovery region, corresponding to a local significance of $0.9\sigma$; this result is included as the last row of Table~\ref{tab:UL}.

\begin{figure}[htbp]
\centering
\begin{subfigure}{0.49\textwidth}
\centering
\includegraphics[width=\textwidth]{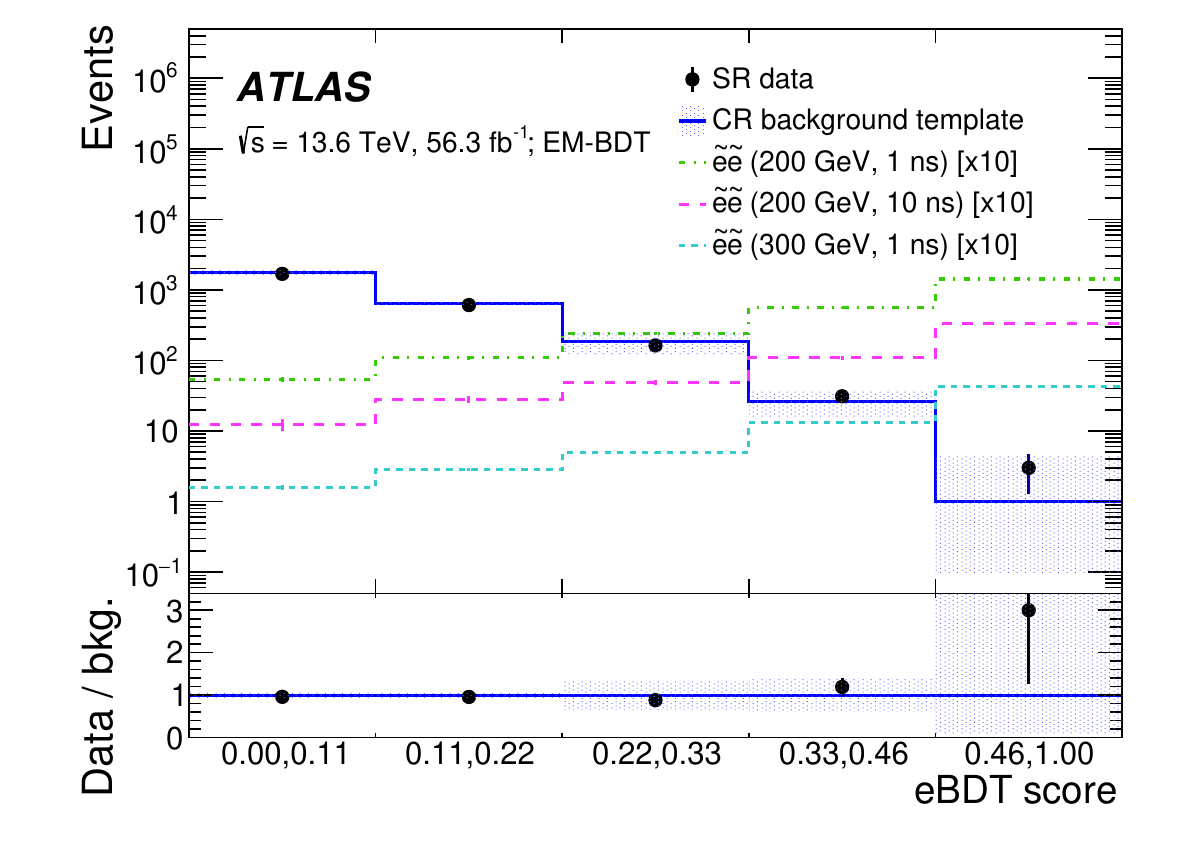}
\caption{}
\end{subfigure}
\begin{subfigure}{0.49\textwidth}
\centering
\includegraphics[width=\textwidth]{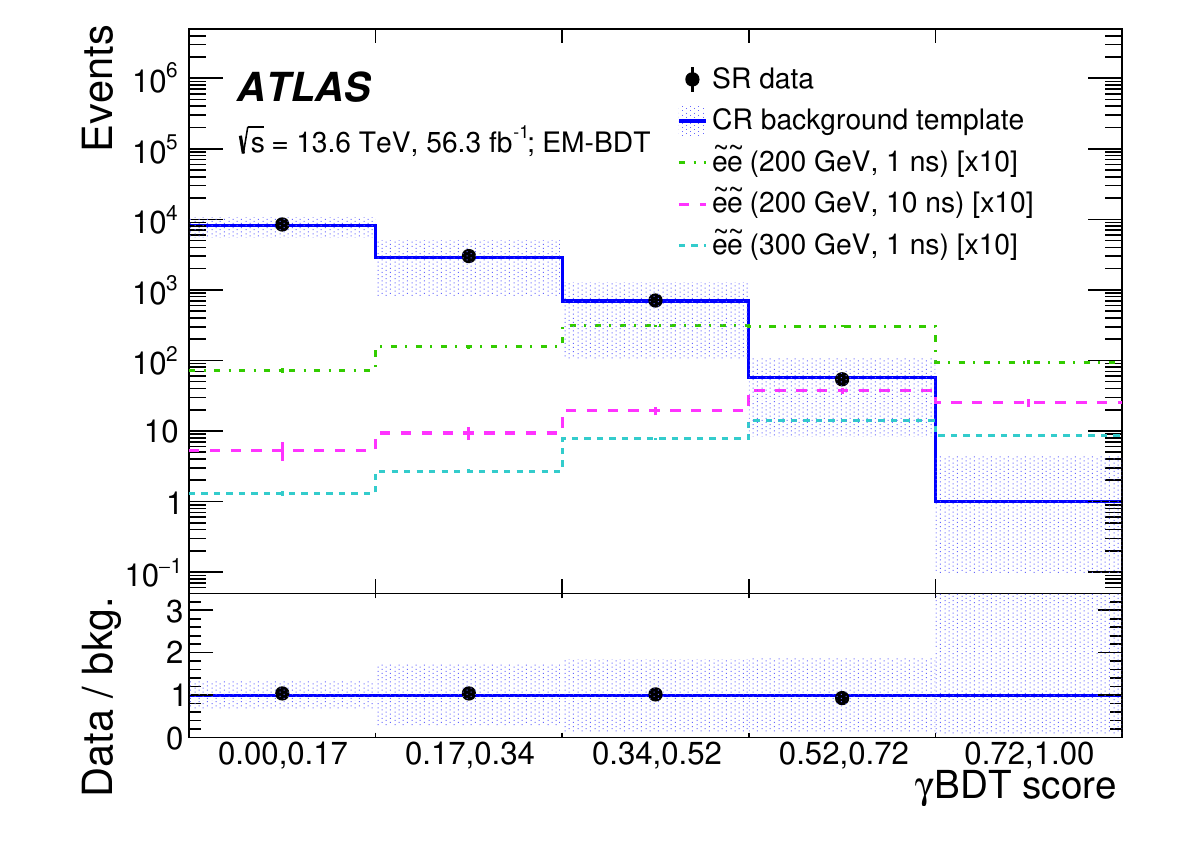}
\caption{}
\end{subfigure}
\caption{BDT score distributions for the (a) $e$BDT and (b) $\gamma$BDT comparing the expected yields from the control region to the observed data.
Selectron pair-production models with a mass of 300~\GeV and lifetime of 1~ns, and mass of 200~\GeV with lifetime of 1~ns or 10~ns are overlaid with 10 times the expected yield.
The lower panel is the ratio of the observed data over the expected background.  }
\label{fig:bdt_score}
\end{figure}

The combined ABCD and EM-BDT results are used to set exclusion contours
on the GMSB stau and selectron models, in the two-dimensional
plane of slepton lifetime versus slepton mass.
Only the ABCD results are used for the smuon and dark sector models.
The combination is obtained by simultaneously fitting all included regions, using a single fit parameter to describe the signal strength.
Thus the 13.6~\TeV to 13~\TeV cross-section ratio is taken from the theoretical prediction.
To improve the interpolation between generated signal models with different lifetimes, a reweighting procedure is used to estimate the signal acceptances for lifetimes that were not included in the MC sample described in Section~\ref{sec:datamc} and corresponding uncertainties in the signal yields are included in the fit.
A 20\% signal normalization uncertainty was assessed for reweighting to lifetimes up to 100 ns, increasing to 50\% for reweighting to even higher lifetimes, where the uncertainties are based on comparing reweighted samples to simulated samples with the same lifetime.

The 95\% CL exclusion
contours for long-lived selectrons, smuons, and staus
are shown in Figure~\ref{fig:slepton_limits}.
Results from the prior Run 2 analysis~\cite{ATLAS:2020wjh} are also included.
The excluded regions from the current analysis extend to significantly higher slepton masses and much higher slepton lifetimes.
For example, selectrons with a mass of 150~\GeV are excluded for lifetime
values ranging from 4~ps to 60~ns, while selectrons, smuons, and staus with a lifetime of 0.3~ns are excluded up to
740~\GeV, 830~\GeV, and 440~\GeV, respectively. %
The improvement in sensitivity for long lifetimes for selectrons with lower masses is due to the EM-BDT and new \ac{LRT}-based triggers.
For smuons, in addition to the \ac{LRT}-based triggers, the improved sensitivity is due to better trigger acceptance for muons in the forward region.
The sensitivity to stau models benefits from all of the improvements above and especially from the new LRT triggers since the leptons tend to have lower \pt.
This is illustrated in Figure~\ref{fig:slepton_limit_breakdown}, which overlays the exclusion contours for the ABCD high-\pt, ABCD LRT, and EM-BDT SRs.
The difference between the expected and observed exclusion contours is due to the observed events in the $ee$ LRT and EM-BDT SRs.
Thus the combined observed limit, including the observed events, can be worse than the standalone high-\pt limit where no events were observed.
The difference between the expected and observed exclusion contours for the stau model comes from the observed event in the ABCD LRT region as the EM-BDT sensitivity is sub-dominant for the mixed flavor final-state from the staus.
All models benefit at high masses from the additional Run 3 data.

The 95\% CL exclusion contours for the dark sector model are shown in Figure~\ref{fig:chargino_limits}.
For a lifetime of $\tau=0.3$~ns, mass differences down to $\Delta m=16$~\GeV\ are excluded for a chargino mass of 100~\GeV, and  charginos with masses up to 900~\GeV are excluded for large values of $\Delta m$.
For a mass difference of $\Delta m=40$~\GeV, charginos with masses up to 380~\GeV are excluded and chargino lifetimes of $0.006<\tau<10$~ns for a chargino mass of 100~\GeV are excluded.
Similar to the exclusion contour overlays for the slepton models, Figure~\ref{fig:chargino_limit_breakdown} illustrates the sensitivity to the chargino models for the ABCD high-\pt SRs at larger masses and sensitivity of the ABCD LRT SRs to the theoretically-preferred small-mass-splittings between 10--40~\GeV~\cite{charginos2}.

\begin{figure}[htbp]
\centering
\begin{subfigure}{0.49\textwidth}
\centering
\includegraphics[width=\textwidth]{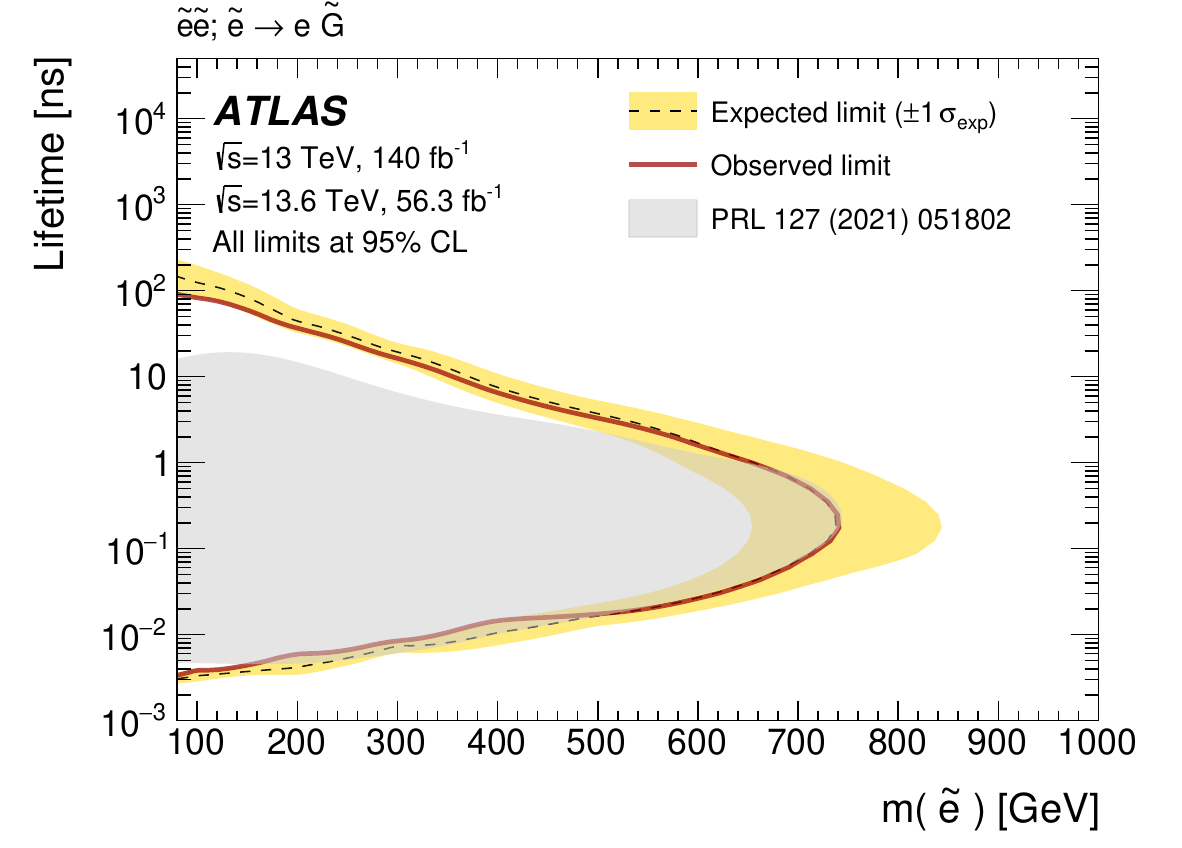}
\caption{}
\end{subfigure}
\begin{subfigure}{0.49\textwidth}
\centering
\includegraphics[width=\textwidth]{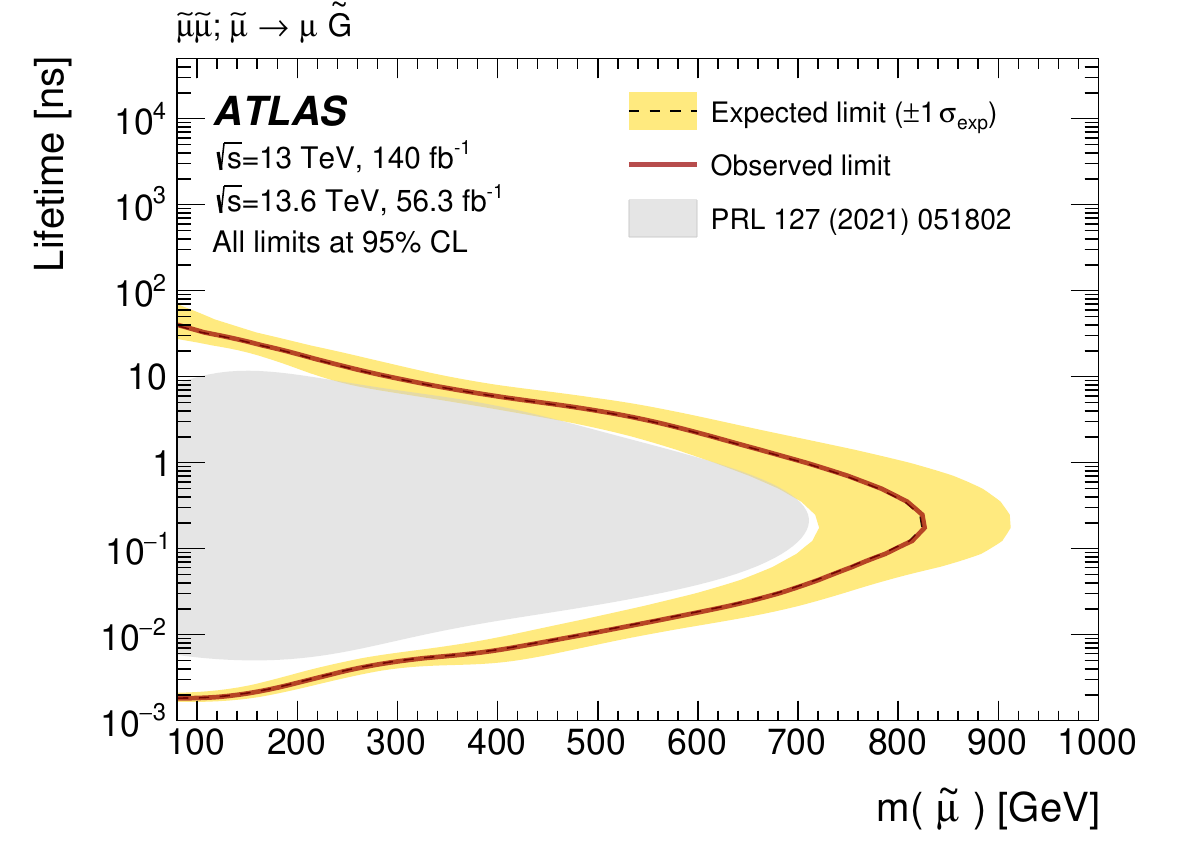}
\caption{}
\end{subfigure}
\begin{subfigure}{0.49\textwidth}
\centering
\includegraphics[width=\textwidth]{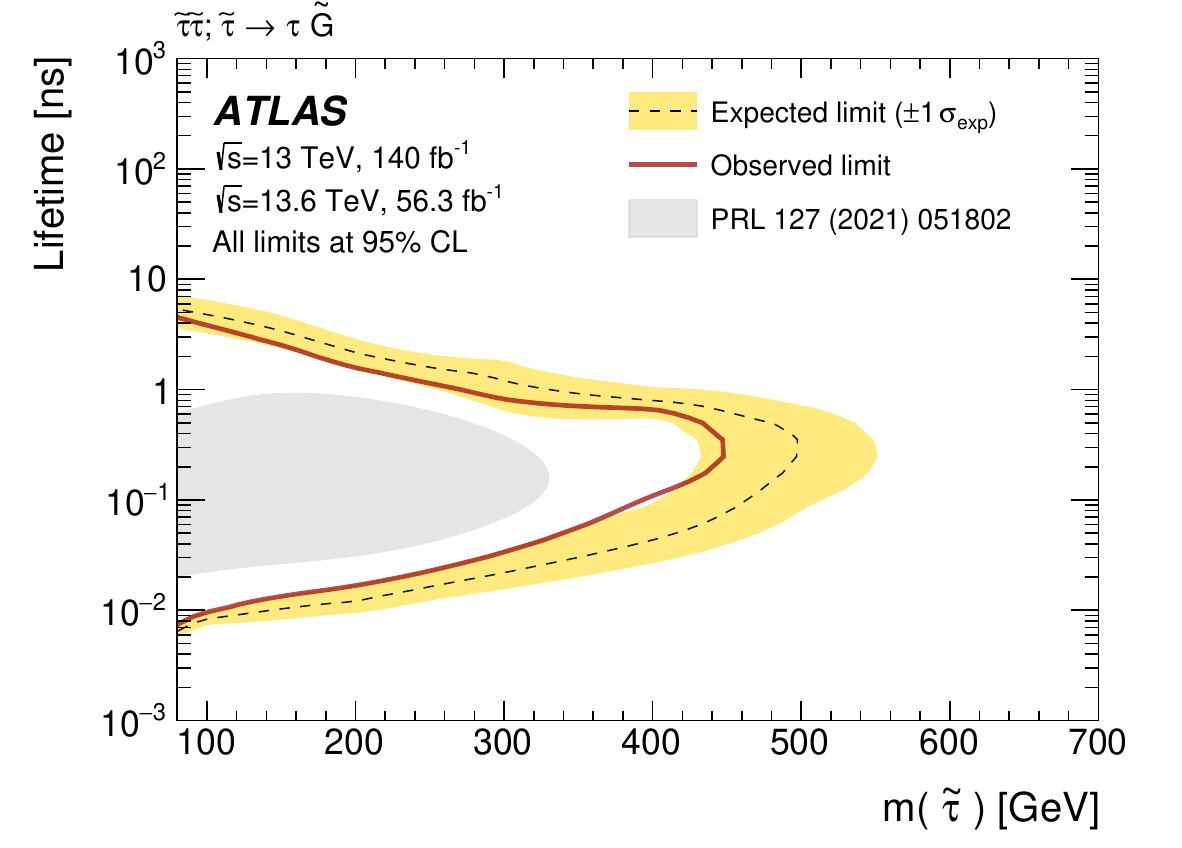}
\caption{}
\end{subfigure}
\caption{Expected and observed 95\% CL exclusion contours for the (a) selectron, (b) smuon, (c) and stau models.
The selectron and stau limits are computed using both the ABCD and EM-BDT regions, while the smuon limits are from the ABCD regions only.
The dashed line indicates the expected limit and the surrounding band shows the $1\sigma$ variation of the expected limit due to uncertainties in the background prediction and experimental uncertainties in the signal modeling.
The gray shaded area illustrates the observed exclusion obtained in the
prior ATLAS Run 2 analysis~\cite{ATLAS:2020wjh}.
}
\label{fig:slepton_limits}
\end{figure}

\begin{figure}[htbp]
\centering
\begin{subfigure}{0.49\textwidth}
\centering
\includegraphics[width=\textwidth]{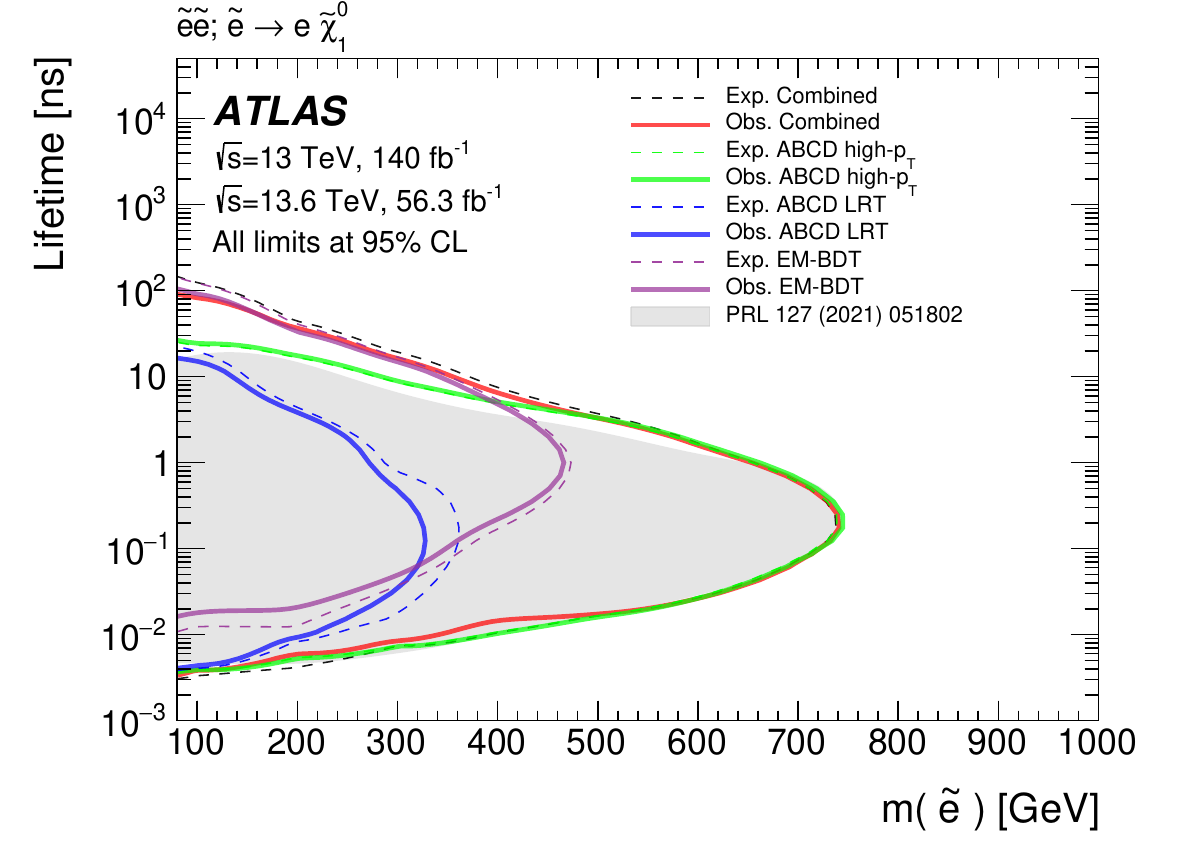}
\caption{}
\end{subfigure}
\begin{subfigure}{0.49\textwidth}
\centering
\includegraphics[width=\textwidth]{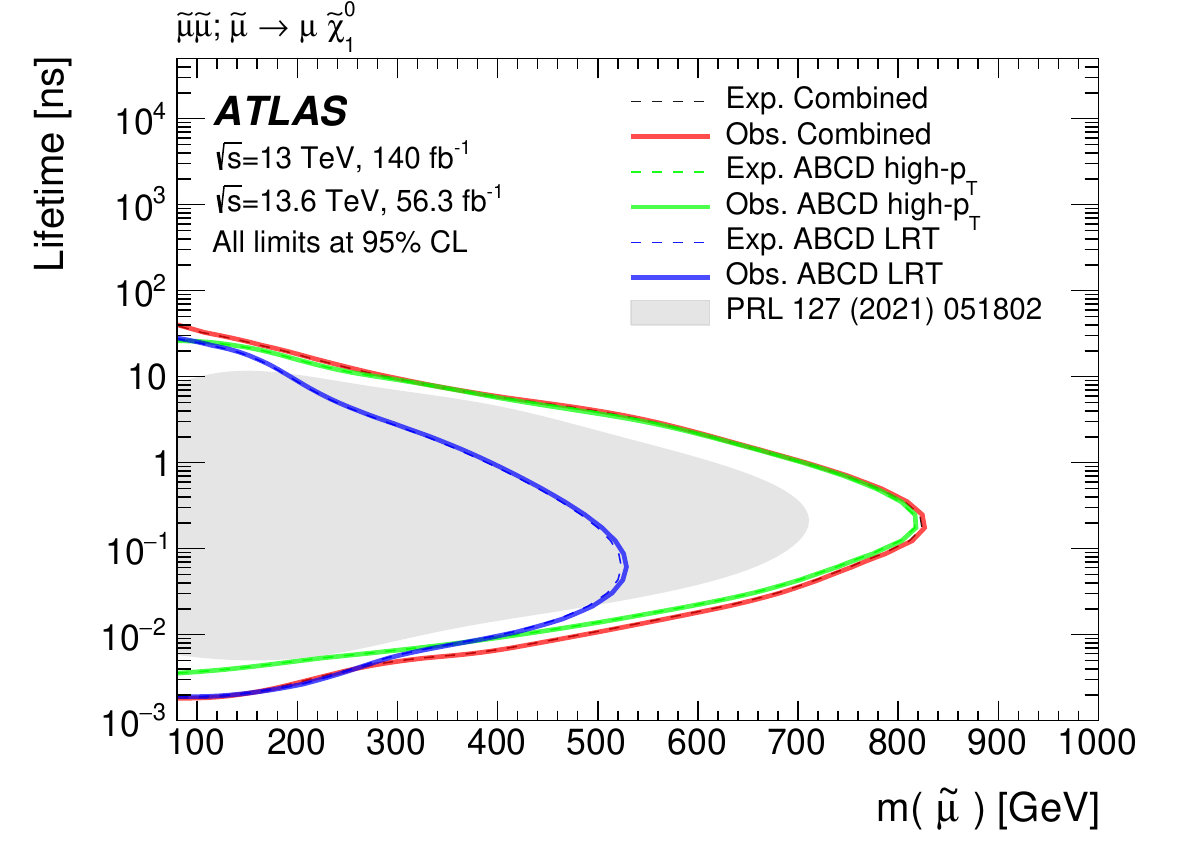}
\caption{}
\end{subfigure}
\begin{subfigure}{0.49\textwidth}
\centering
\includegraphics[width=\textwidth]{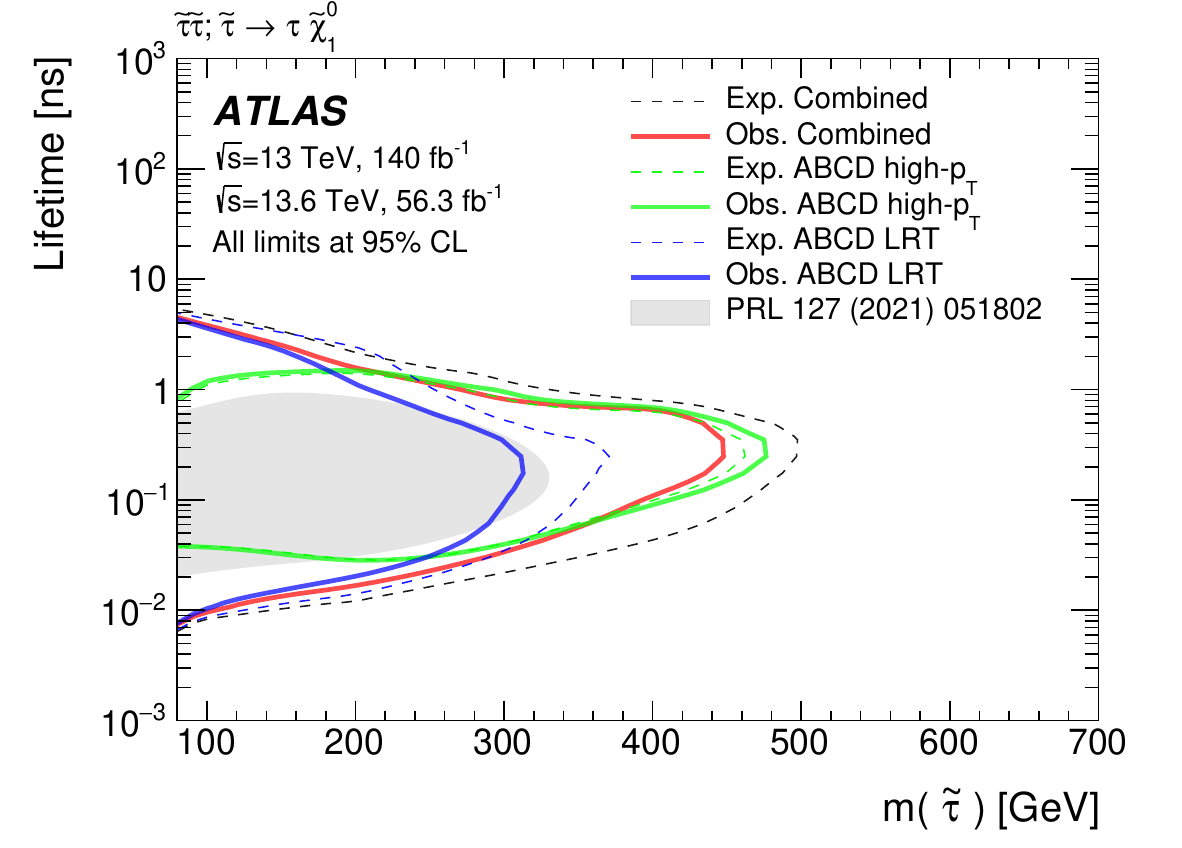}
\caption{}
\end{subfigure}
\caption{Comparison of expected and observed 95\% CL exclusion contours for the (a) selectron, (b) smuon, and (c) stau models for subsets of the analysis regions: ABCD high-\pt, ABCD LRT, and EM-BDT.
For the stau model, the EM-BDT SRs do not exclude any points by themselves.
The dashed lines indicate the expected limits.
The gray shaded area illustrates the observed exclusion obtained in the
prior ATLAS Run 2 analysis~\cite{ATLAS:2020wjh}. }
\label{fig:slepton_limit_breakdown}
\end{figure}

\begin{figure}[htbp]
\centering
\begin{subfigure}{0.49\textwidth}
\centering
\includegraphics[width=\textwidth]{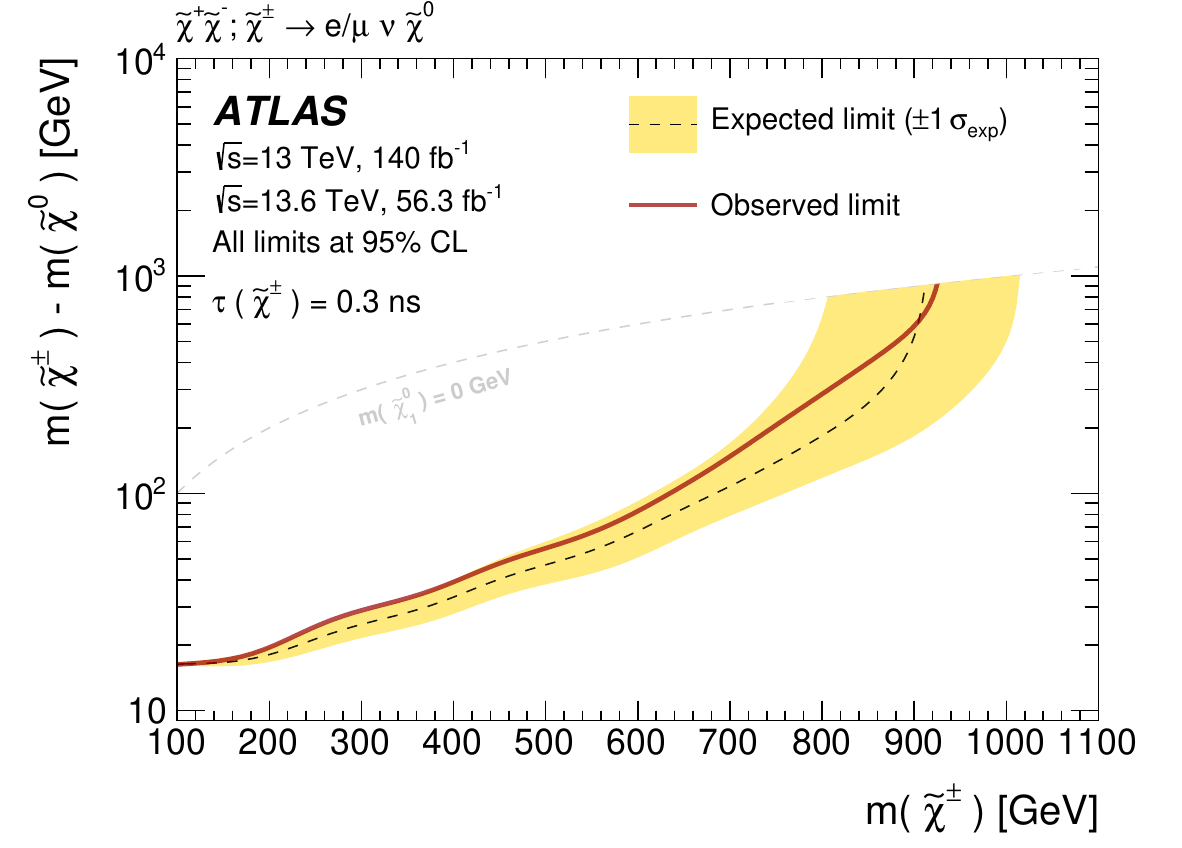}
\caption{}
\end{subfigure}
\begin{subfigure}{0.49\textwidth}
\centering
\includegraphics[width=\textwidth]{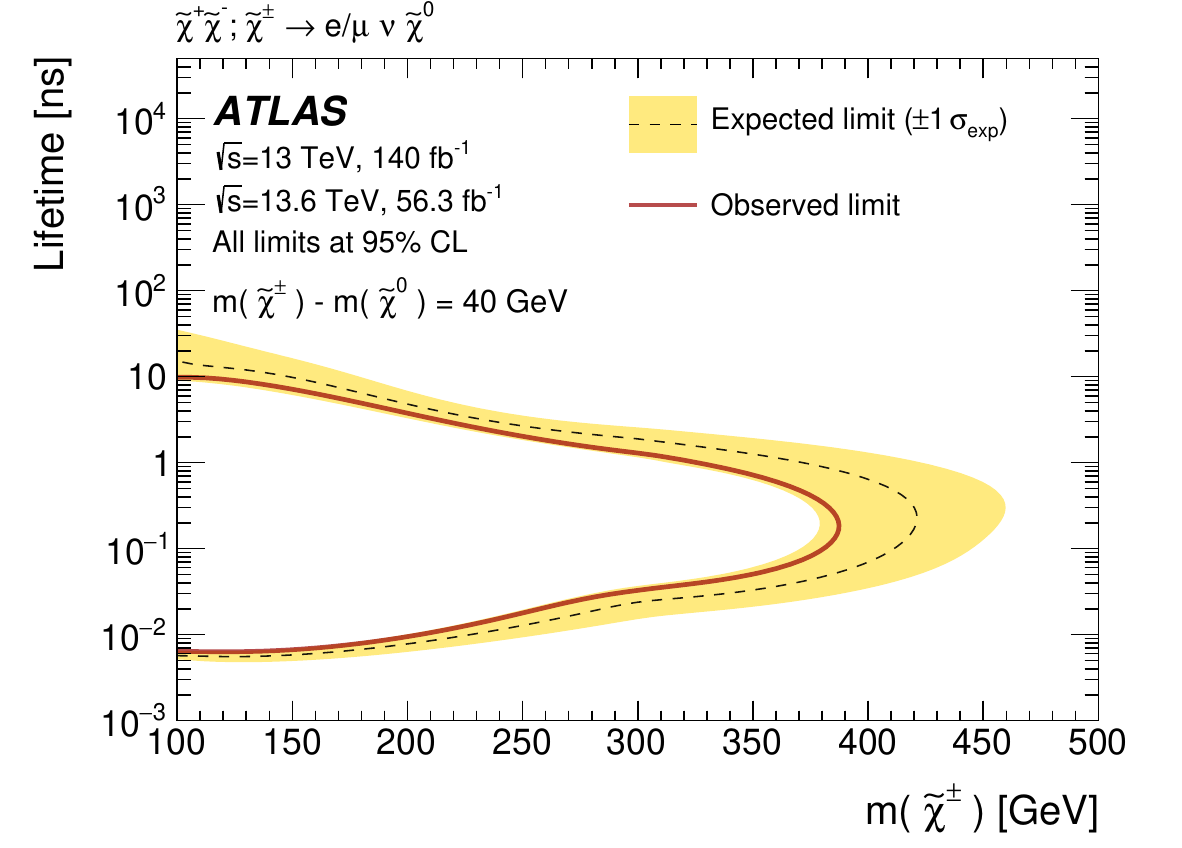}
\caption{}
\end{subfigure}
\caption{Expected and observed 95\% CL exclusion contours for the dark sector model with a (a) fixed lifetime and (b) fixed mass splitting.
The dashed line indicates the expected limit and the surrounding band shows the $1\sigma$ variation of the expected limit due to uncertainties in the background prediction and experimental uncertainties in the signal modeling.
}
\label{fig:chargino_limits}
\end{figure}

\begin{figure}[htbp]
\centering
\begin{subfigure}{0.49\textwidth}
\centering
\includegraphics[width=\textwidth]{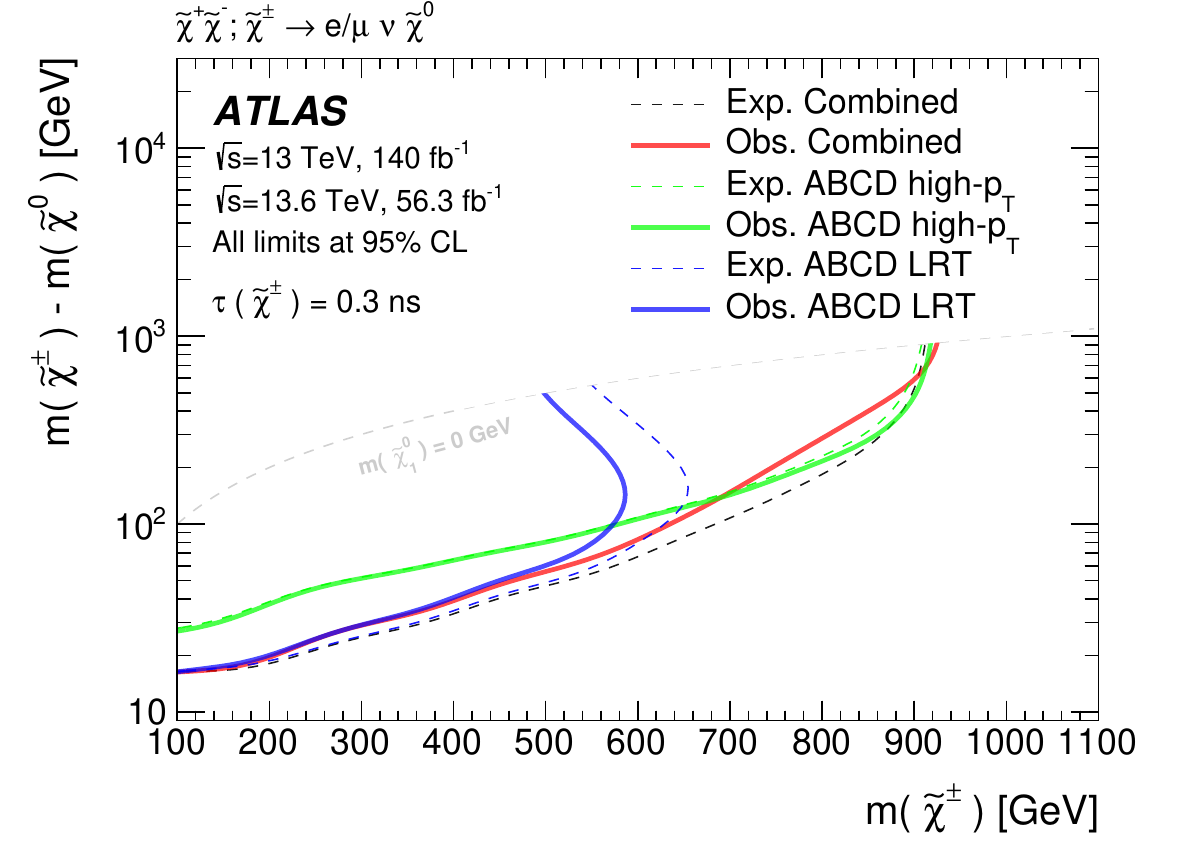}
\caption{}
\end{subfigure}
\begin{subfigure}{0.49\textwidth}
\centering
\includegraphics[width=\textwidth]{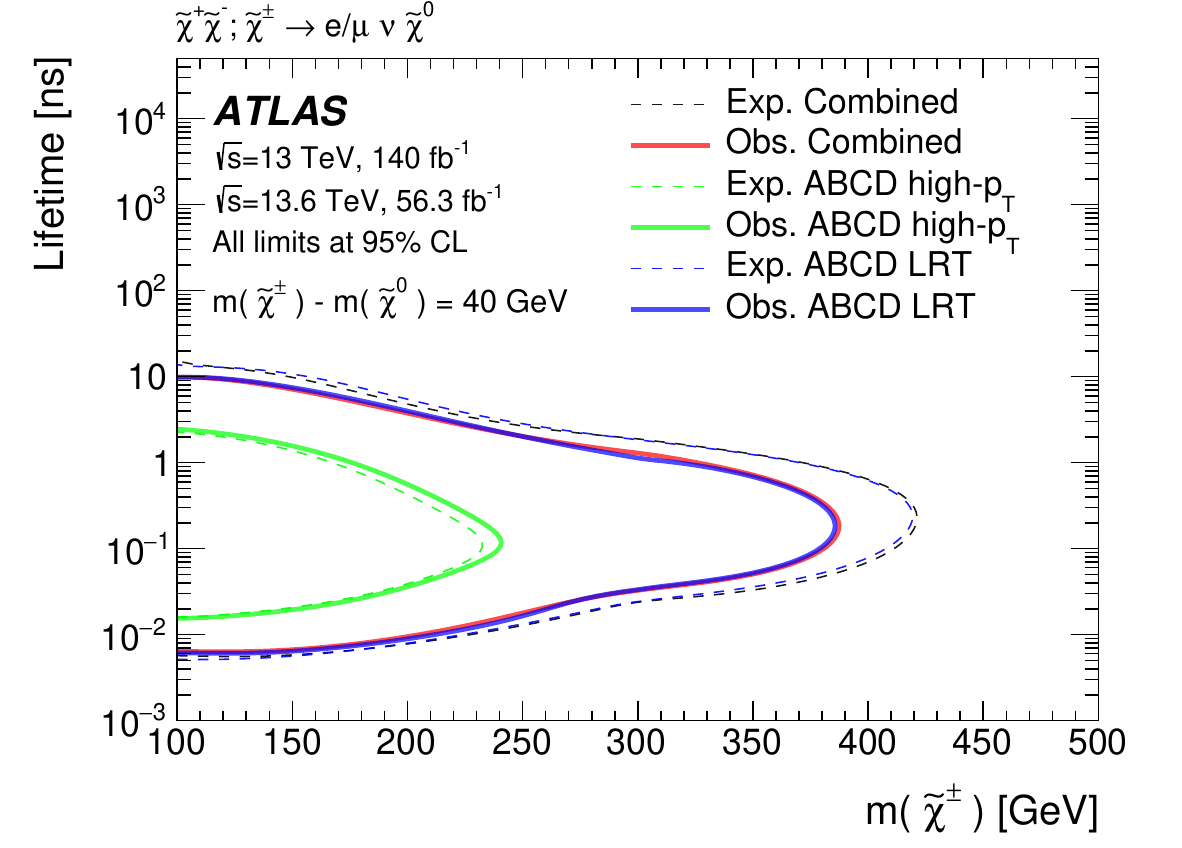}
\caption{}
\end{subfigure}
\caption{Comparison of expected and observed 95\% CL exclusion contours for the dark sector model with a (a) fixed-lifetime and (b) fixed-mass-splitting for subsets of the analysis regions: ABCD high-\pt and ABCD LRT.
The dashed lines indicate the expected limits. }
\label{fig:chargino_limit_breakdown}
\end{figure}


%

%
%
%

\FloatBarrier
\section{Conclusion}
\label{sec:conclusion}


A search for \ac{BSM} physics using 56.3~\ifb of proton--proton collision data collected by the ATLAS detector at the LHC during 2022--2023 with $\sqrt{s} = 13.6$~TeV, combined with the full Run 2 dataset in some regions, was presented.
The analysis focuses on the signature of displaced electrons or muons, consistent with the decay of a long-lived BSM particle.
The event selection exploits new Run 3 triggers that employ LRT for sensitivity to lower momentum final state particles.
Two orthogonal approaches are used, one focusing on dilepton events and another using LAr measurements and multivariate techniques for expanded sensitivity to single-electron and photon-reconstructed events.

The observed event yields in the various signal regions are consistent with the SM background expectations.
The results are therefore used to set model-independent limits on the production of displaced electrons
and muons.
The combined analysis results are interpreted with two models.
The first is a GMSB model with  pair-produced long-lived sleptons that decay into the lepton of the same flavor and a nearly massless gravitino, providing exclusion contours that extend significantly beyond previous results.
For example,
95\% CL exclusions are obtained for selectrons with a mass of 150~\GeV and lifetime values ranging from 4~ps to 60~ns, and selectrons, smuons, and staus with a 0.3~ns lifetime are excluded up to masses of
740~\GeV, 830~\GeV, and 440~\GeV, respectively.
The second model is pair-produced dark charginos decaying into a $W$ boson and neutralino, where small mass splittings of 10--40~\GeV result in a dark matter relic density consistent with observations.
Similarly, 95\% CL are set for charginos with masses up to 380~\GeV for a mass
difference with the neutralino of 40~\GeV, and mass differences down to 17~\GeV are excluded for
dark charginos with a 100~\GeV mass.


%
%

%
%
%
%
%
\section*{Acknowledgments}
%

%

%
%

%
%

We thank CERN for the very successful operation of the LHC and its injectors, as well as the support staff at
CERN and at our institutions worldwide without whom ATLAS could not be operated efficiently.

The crucial computing support from all WLCG partners is acknowledged gratefully, in particular from CERN, the ATLAS Tier-1 facilities at TRIUMF/SFU (Canada), NDGF (Denmark, Norway, Sweden), CC-IN2P3 (France), KIT/GridKA (Germany), INFN-CNAF (Italy), NL-T1 (Netherlands), PIC (Spain), RAL (UK) and BNL (USA), the Tier-2 facilities worldwide and large non-WLCG resource providers. Major contributors of computing resources are listed in Ref.~\cite{ATL-SOFT-PUB-2025-001}.

We gratefully acknowledge the support of ANPCyT, Argentina; YerPhI, Armenia; ARC, Australia; BMWFW and FWF, Austria; ANAS, Azerbaijan; CNPq and FAPESP, Brazil; NSERC, NRC and CFI, Canada; CERN; ANID, Chile; CAS, MOST and NSFC, China; Minciencias, Colombia; MEYS CR, Czech Republic; DNRF and DNSRC, Denmark; IN2P3-CNRS and CEA-DRF/IRFU, France; SRNSFG, Georgia; BMBF, HGF and MPG, Germany; GSRI, Greece; RGC and Hong Kong SAR, China; ISF and Benoziyo Center, Israel; INFN, Italy; MEXT and JSPS, Japan; CNRST, Morocco; NWO, Netherlands; RCN, Norway; MNiSW, Poland; FCT, Portugal; MNE/IFA, Romania; MSTDI, Serbia; MSSR, Slovakia; ARIS and MVZI, Slovenia; DSI/NRF, South Africa; MICIU/AEI, Spain; SRC and Wallenberg Foundation, Sweden; SERI, SNSF and Cantons of Bern and Geneva, Switzerland; NSTC, Taipei; TENMAK, T\"urkiye; STFC/UKRI, United Kingdom; DOE and NSF, United States of America.

Individual groups and members have received support from BCKDF, CANARIE, CRC and DRAC, Canada; CERN-CZ, FORTE and PRIMUS, Czech Republic; COST, ERC, ERDF, Horizon 2020, ICSC-NextGenerationEU and Marie Sk{\l}odowska-Curie Actions, European Union; Investissements d'Avenir Labex, Investissements d'Avenir Idex and ANR, France; DFG and AvH Foundation, Germany; Herakleitos, Thales and Aristeia programmes co-financed by EU-ESF and the Greek NSRF, Greece; BSF-NSF and MINERVA, Israel; NCN and NAWA, Poland; La Caixa Banking Foundation, CERCA Programme Generalitat de Catalunya and PROMETEO and GenT Programmes Generalitat Valenciana, Spain; G\"{o}ran Gustafssons Stiftelse, Sweden; The Royal Society and Leverhulme Trust, United Kingdom.

In addition, individual members wish to acknowledge support from Armenia: Yerevan Physics Institute (FAPERJ); CERN: European Organization for Nuclear Research (CERN PJAS); Chile: Agencia Nacional de Investigaci\'on y Desarrollo (FONDECYT 1230812, FONDECYT 1230987, FONDECYT 1240864); China: Chinese Ministry of Science and Technology (MOST-2023YFA1605700, MOST-2023YFA1609300), National Natural Science Foundation of China (NSFC - 12175119, NSFC 12275265, NSFC-12075060); Czech Republic: Czech Science Foundation (GACR - 24-11373S), Ministry of Education Youth and Sports (ERC-CZ-LL2327, FORTE CZ.02.01.01/00/22\_008/0004632), PRIMUS Research Programme (PRIMUS/21/SCI/017); EU: H2020 European Research Council (ERC - 101002463); European Union: European Research Council (ERC - 948254, ERC 101089007), Horizon 2020 Framework Programme (MUCCA - CHIST-ERA-19-XAI-00), European Union, Future Artificial Intelligence Research (FAIR-NextGenerationEU PE00000013), Italian Center for High Performance Computing, Big Data and Quantum Computing (ICSC, NextGenerationEU); France: Agence Nationale de la Recherche (ANR-20-CE31-0013, ANR-21-CE31-0013, ANR-21-CE31-0022, ANR-22-EDIR-0002), Investissements d'Avenir Labex (ANR-11-LABX-0012); Germany: Baden-Württemberg Stiftung (BW Stiftung-Postdoc Eliteprogramme), Deutsche Forschungsgemeinschaft (DFG - 469666862, DFG - CR 312/5-2); Italy: Istituto Nazionale di Fisica Nucleare (ICSC, NextGenerationEU), Ministero dell'Università e della Ricerca (NextGenEU I53D23000820006 M4C2.1.1); Japan: Japan Society for the Promotion of Science (JSPS KAKENHI JP22H01227, JSPS KAKENHI JP22H04944, JSPS KAKENHI JP22KK0227, JSPS KAKENHI JP23KK0245); Netherlands: Netherlands Organisation for Scientific Research (NWO Veni 2020 - VI.Veni.202.179); Norway: Research Council of Norway (RCN-314472); Poland: Ministry of Science and Higher Education (IDUB AGH, POB8, D4 no 9722), Polish National Agency for Academic Exchange (PPN/PPO/2020/1/00002/U/00001), Polish National Science Centre (NCN 2021/42/E/ST2/00350, NCN OPUS nr 2022/47/B/ST2/03059, NCN UMO-2019/34/E/ST2/00393, NCN \& H2020 MSCA 945339, UMO-2020/37/B/ST2/01043, UMO-2021/40/C/ST2/00187, UMO-2022/47/O/ST2/00148, UMO-2023/49/B/ST2/04085, UMO-2023/51/B/ST2/00920); Portugal: Foundation for Science and Technology (FCT); Slovenia: Slovenian Research Agency (ARIS grant J1-3010); Spain: Generalitat Valenciana (Artemisa, FEDER, IDIFEDER/2018/048), Ministry of Science and Innovation (MCIN \& NextGenEU PCI2022-135018-2, MICIN \& FEDER PID2021-125273NB, RYC2019-028510-I, RYC2020-030254-I, RYC2021-031273-I, RYC2022-038164-I); Sweden: Carl Trygger Foundation (Carl Trygger Foundation CTS 22:2312), Swedish Research Council (Swedish Research Council 2023-04654, VR 2018-00482, VR 2021-03651, VR 2022-03845, VR 2022-04683, VR 2023-03403), Knut and Alice Wallenberg Foundation (KAW 2018.0458, KAW 2019.0447, KAW 2022.0358); Switzerland: Swiss National Science Foundation (SNSF - PCEFP2\_194658); United Kingdom: Leverhulme Trust (Leverhulme Trust RPG-2020-004), Royal Society (NIF-R1-231091); United States of America: U.S. Department of Energy (ECA DE-AC02-76SF00515), Neubauer Family Foundation.

%
%


%
%
%
%
%
%

%
%
%
\printbibliography
\clearpage
 
\begin{flushleft}
\hypersetup{urlcolor=black}
{\Large The ATLAS Collaboration}

\bigskip

\AtlasOrcid[0000-0002-6665-4934]{G.~Aad}$^\textrm{\scriptsize 105}$,
\AtlasOrcid[0000-0001-7616-1554]{E.~Aakvaag}$^\textrm{\scriptsize 17}$,
\AtlasOrcid[0000-0002-5888-2734]{B.~Abbott}$^\textrm{\scriptsize 124}$,
\AtlasOrcid[0000-0002-0287-5869]{S.~Abdelhameed}$^\textrm{\scriptsize 120a}$,
\AtlasOrcid[0000-0002-1002-1652]{K.~Abeling}$^\textrm{\scriptsize 57}$,
\AtlasOrcid[0000-0001-5763-2760]{N.J.~Abicht}$^\textrm{\scriptsize 51}$,
\AtlasOrcid[0000-0002-8496-9294]{S.H.~Abidi}$^\textrm{\scriptsize 30}$,
\AtlasOrcid[0009-0003-6578-220X]{M.~Aboelela}$^\textrm{\scriptsize 46}$,
\AtlasOrcid[0000-0002-9987-2292]{A.~Aboulhorma}$^\textrm{\scriptsize 36e}$,
\AtlasOrcid[0000-0001-5329-6640]{H.~Abramowicz}$^\textrm{\scriptsize 156}$
% ...
\bigskip
\\

% ...
$^{17}$Department for Physics and Technology, University of Bergen, Bergen; Norway.\\
% ...
$^{30}$Physics Department, Brookhaven National Laboratory, Upton NY; United States of America.\\
% ...
$^{36}$$^{(a)}$Facult\'e des Sciences Ain Chock, Universit\'e Hassan II de Casablanca;$^{(b)}$Facult\'{e} des Sciences, Universit\'{e} Ibn-Tofail, K\'{e}nitra;$^{(c)}$Facult\'e des Sciences Semlalia, Universit\'e Cadi Ayyad, LPHEA-Marrakech;$^{(d)}$LPMR, Facult\'e des Sciences, Universit\'e Mohamed Premier, Oujda;$^{(e)}$Facult\'e des sciences, Universit\'e Mohammed V, Rabat;$^{(f)}$Institute of Applied Physics, Mohammed VI Polytechnic University, Ben Guerir; Morocco.\\
% ...
$^{46}$Physics Department, Southern Methodist University, Dallas TX; United States of America.\\
% ...
$^{51}$Fakult\"{a}t Physik , Technische Universit{\"a}t Dortmund, Dortmund; Germany.\\
% ...
$^{57}$II. Physikalisches Institut, Georg-August-Universit\"{a}t G\"ottingen, G\"ottingen; Germany.\\
% ...
$^{105}$CPPM, Aix-Marseille Universit\'e, CNRS/IN2P3, Marseille; France.\\
% ...
$^{120}$$^{(a)}$New York University Abu Dhabi, Abu Dhabi;$^{(b)}$United Arab Emirates University, Al Ain; United Arab Emirates.\\
% ...
$^{124}$Homer L. Dodge Department of Physics and Astronomy, University of Oklahoma, Norman OK; United States of America.\\
% ...
$^{156}$Raymond and Beverly Sackler School of Physics and Astronomy, Tel Aviv University, Tel Aviv; Israel.\\
% ...
$^{a}$ Also Affiliated with an institute covered by a cooperation agreement with CERN.\\
$^{b}$ Also at An-Najah National University, Nablus; Palestine.\\
$^{c}$ Also at Borough of Manhattan Community College, City University of New York, New York NY; United States of America.\\
$^{d}$ Also at Center for High Energy Physics, Peking University; China.\\
$^{e}$ Also at Center for Interdisciplinary Research and Innovation (CIRI-AUTH), Thessaloniki; Greece.\\
$^{f}$ Also at CERN, Geneva; Switzerland.\\
% ...

\end{flushleft}


%

%
%

%

%
%
%
%

\end{document}